%% file: main.tex
\documentclass[journal]{IEEEtran}

\usepackage{graphicx}
\usepackage[toc]{glossaries}

\usepackage{float}
\restylefloat{table}

\usepackage[hidelinks]{hyperref}
\hypersetup{
    colorlinks=true,
    linkcolor=blue,
    filecolor=magenta,
    urlcolor=cyan,
}

\usepackage{amsmath}
\usepackage{amssymb}
\usepackage{eucal}
\usepackage{booktabs}
\usepackage{caption,subcaption}
\usepackage{changes}
\usepackage{pifont}
\usepackage{rotating}
\usepackage{framed}
\usepackage{type1cm}
\usepackage{lettrine}
\usepackage{tcolorbox}
\usepackage{nicefrac,xfrac}
\usepackage{lipsum,lineno}
\usepackage{makecell}
\usepackage{xspace}
\usepackage{scalefnt}

\usepackage{float}
\restylefloat{table}

\usepackage{tikz}
\usetikzlibrary{backgrounds}

\let\labelindent\relax
\usepackage{enumitem}

\newcommand{\figref}[1]{Fig.~\ref{#1}}
\newcommand{\tableref}[1]{Table~\ref{#1}}
\renewcommand{\eqref}[1]{(\ref{#1})}
\newcommand{\secref}[1]{Section~\ref{#1}}

\newcommand{\loss}{data-adaptive loss function}
\newcommand{\losspl}{\loss{}s}

\newcommand{\ie}{\textit{i.e.}\xspace}

\newcommand{\mycomment}[1]{}

\usepackage{adjustbox}
\usepackage{array}
\usepackage{multirow}
\newcolumntype{L}[1]{>{\raggedright\let\newline\\\arraybackslash\hspace{0pt}}m{#1}}
\newcolumntype{C}[1]{>{\centering\let\newline\\\arraybackslash\hspace{0pt}}m{#1}}

\newcolumntype{R}[2]{%
    >{\adjustbox{angle=#1,lap=\width-(#2)}\bgroup}%
    c%
    <{\egroup}%
}
\newcommand*\rotso{\multicolumn{1}{R{20}{1em}}}

\include{glossaries}

\begin{document}
\bstctlcite{IEEEexample:BSTcontrol}

\title{A Data-Adaptive Loss Function for Incomplete Data and Incremental Learning in Semantic Image Segmentation}

\author{Minh H. Vu*,
		Gabriella Norman*,
        Tufve Nyholm,		
        and Tommy L\"{o}fstedt
\thanks{Copyright (c) 2021 IEEE. Personal use of this material is permitted. However, permission to use this material for any other purposes must be obtained from the IEEE by sending a request to pubs-permissions@ieee.org.}        
\thanks{The computations were enabled by resources provided by the \gls{snic} at the \gls{hpc2n} in Ume{\aa}, Sweden, partially funded by the Swedish Research Council through grant agreement no. 2018-05973.
We are grateful for the financial support obtained from the Cancer Research Fund in Northern Sweden, Karin and Krister Olsson, Ume\aa{} University, The V\"{a}sterbotten regional county, and Vinnova, the Swedish innovation agency. This research was in part supported by grants from the Cancer Research Foundation in Northern Sweden (AMP 20-1027, LP 18-2182).}
\thanks{*M. H. Vu and G. Norman contribute equally to the work.}
\thanks{M. H. Vu, G. Norman, and T. Nyholm are with the Department of Radiation Sciences, Radiation Physics, Ume{\aa} University, Ume{\aa}, Sweden.}
\thanks{T. L\"{o}fstedt is with Department of Computing Science, Ume{\aa} University, Ume{\aa}, Sweden. E-mail: tommy@cs.umu.se.}
}

\maketitle              

\begin{abstract}

In the last years, deep learning has dramatically improved the performances in a variety of medical image analysis applications. Among different types of deep learning models, convolutional neural networks have been among the most successful and they have been used in many applications in medical imaging. 

Training deep convolutional neural networks often requires large amounts of image data to generalize well to new unseen images. It is often time-consuming and expensive to collect large amounts of data in the medical image domain due to expensive imaging systems, and the need for experts to manually make ground truth annotations. A potential problem arises if new structures are added when a decision support system is already deployed and in use. Since the field of radiation therapy is constantly developing, the new structures would also have to be covered by the decision support system.

In the present work, we propose a novel loss function, that adapts to the available data in order to utilize all available data, even when some have missing annotations. We demonstrate that the proposed loss function also works well in an incremental learning setting, where it can automatically incorporate new structures as they appear. Experiments on a large in-house data set show that the proposed method performs on par with baseline models, while greatly reducing the training time.

\end{abstract}


\section{Introduction}

\lettrine{C}{ancer} is the second leading cause of death globally and accounted for an estimated 10 million deaths in 2020~\cite{Sung2021}. In Sweden, more than 60\,000 patients are diagnosed each year, and about half of them undergo radiation therapy~\cite{JONSSON201445}. Before the radiation therapy can begin, an oncologist manually marks, or delineate, the regions in the body that should be treated (target volume) and the regions that are particularly important to avoid. The delineations are then used to generate a dose plan that gives a sufficient dose of radiation to the target areas, and a small dose (as small as possible) to sensitive organs, called \glspl{oar}. For that goal to be achieved, the delineations must be correct. Hence, delineation is an essential part of the treatment planning process, but a time-consuming and monotonic manual task for radiation oncologists. Therefore, decision support systems that automate the delineation process would be beneficial in order to reduce the amount of time spent on the challenging task of manually delineating target volumes and organs~\cite{risk_organ_seg_zhikai, multi_organ_sulaiman}. 
In addition, the automatic delineation would also make it possible to delineate more organs at risk. This would in time lead to a better understanding of the relation to dose to certain volumes and side effects of the treatment.

Recently, \gls{dl} methods, and in particular deep \glspl{cnn} have led to breakthroughs in multiple areas of medical imaging. A common application among those is automatic segmentation of organs and \glspl{oar}~\cite{review_DL_shen}, which---if used to automate all or parts of the delineation stage---would reduce the time spent by radiation oncologists manually delineating images. However, deep neural networks, such as deep \glspl{cnn}, require large amounts of data; but data is usually challenging and expensive to collect in the medical image domain due to expensive imaging systems, and the requirement to have experts manually annotate ground truth targets or labels~\cite{review_DL_shen, semi_3_Bai}.

Moreover, since the field of radiation therapy is improving and developing, organs are sometimes proposed to be added as \glspl{oar}~\cite{LADR, penile_bulb}
and therefore new data would be required in order to provide decision support for those newly added \glspl{oar} as well. Two examples of when the clinical practice can change are from \textit{e.g.}~Lee \textit{et al.}~\cite{LADR} who proposed to include the left anterior descending coronary artery region as an \gls{oar} when treating breast cancer and from Roach \textit{et al.}~\cite{penile_bulb} who proposed to include the penile bulb for prostate cancer.

These circumstances can cause projects to spend extensive resources on creating curated data sets, spend numerous hours developing and training deep neural networks, or other \gls{ml} models, to provide decision support, only to have to redo it when a new \gls{oar} has to be included. To add a new \gls{oar} to a decision support system requires adding new data or annotations to the data set and then retrain the \gls{ml} models from their initial configurations using all the data.

A desirable approach would instead be to use the new data created during treatment planning and utilize incremental learning~\cite{incremental}, \ie to automatically incorporate new delineations when they appear. Data created during clinical practice is likely incomplete, \textit{e.g.}~different patients could have different \glspl{oar} delineated, and several delineations may therefore be missing for most, if not all patients.

To approach the problem of missing delineations, we propose a novel loss function that adapts to the available data, allowing incomplete or missing sets of annotations or labels, and allowing new annotations or labels to be added without retraining. The task that was investigated in this work was thus automatic segmentation of target volumes and \glspl{oar}, and the proposed loss function then allows: (i) some patients to provide information regarding some delineations, (ii) other patients to provide information about other delineations---with some target volumes or \glspl{oar} being delineated in most or all patients, and (iii) other \glspl{oar} being delineated only in some patients. Since the data is incomplete to begin with, the hypothesis was that adding new \glspl{oar} during or after having already trained a model would be possible without losing the performance on the already available \glspl{oar}.

The proposed loss is used in conjunction with deep \gls{cnn} models, and works as follows:
Available delineations are predicted by the model and contribute to the loss, while unavailable delineations are still predicted by the network, but are excluded from the loss since there are no ground truth delineations to compare with. In the end, the network will be able to predict all considered delineations. See \secref{sec:loss} for a detailed description of the proposed loss function.

A similar idea was proposed by Liu \textit{et al.}~\cite{Incomp_brain}, but for predicting brain disease prognosis. Their model predicted 16 clinical scores for \gls{mri} data but was trained on data where parts did not have ground truths for all the 16 scores. A label was used to include or exclude the scores from the loss function depending on if they existed or not~\cite{Incomp_brain}. Liu \textit{et al.} worked in the regression setting, while this work was independently developed with the classification task in mind (segmentation is classification of each individual pixel); further, the possibility of adding new clinical scores was never investigated by Liu \textit{et al.}, neither was a comparison to \gls{ssl}.

\gls{ssl} is an approach to \gls{ml} that utilizes both labeled data (supervised) as well as unlabeled data (unsupervised) during training. A common contemporary approach in multi-organ segmentation when training on incomplete data is to use \gls{ssl}~\cite{semi_3_Bai, semi_1_Zhou}. For instance, Zhou \textit{et al.}~\cite{semi_1_Zhou} used 210 labeled cases and 100 unlabeled cases of \gls{ct} scans to train and validate a segmentation \gls{ssl} model. Their model outperformed a fully supervised model by more than 4~\% in terms of dice score. Their \gls{ssl} model has a \textit{teacher model} that is trained on the labeled data and then used to generate \textit{pseudo-labels} for the unlabeled data. It then also has a \textit{student model} that is trained on both the labeled and the pseudo-labeled data.

A similar approach can be applied to the incomplete data problem in our setting, where patients only have ground truth delineations for some of the \glspl{oar}. One teacher model for each delineation would be needed to fill the gaps in the data before training a student model to predict all delineations. However, this approach does not allow adding new \glspl{oar} after a student model has already been trained. In the case that new \glspl{oar} are introduced, a teacher model would have to be trained for the new \glspl{oar} to fill the gaps in the historical data, and then the student model would have to be retrained from its initial configuration on the labeled and pseudo-labeled data.

In this work, we explored the properties of the proposed \loss{} by comparing it to individual (single) models for each label and to the \gls{ssl} approach in the segmentation task. In addition, we also looked at how well a model trained using the \loss{} can adapt to new \glspl{oar} being introduced by mimicking the circumstances in a clinical setting, where new \glspl{oar} are added to an already available decision support system. 

The paper is structured as follows. We introduce the proposed method in \secref{sec:proposed}, and describe the implementation details and training in \secref{sec:details}. Finally, we present our experimental results and discussion in \secref{sec:results} and \secref{sec:discussion}, respectively..

\section{Proposed Method} \label{sec:proposed}

\subsection{Data-Adaptive Loss Function} \label{sec:loss}

The basis for the proposed loss function is a convex combination of the soft \gls{dsc} loss and the \gls{ce} loss. The combination of \gls{dsc} and \gls{ce} have been successful when training segmentation networks, especially if the structures are unbalanced in size, meaning that there is a (large) disparity in the number of pixels between different segmentation maps~\cite{combo_loss_saeid, comb_loss_ken}. The combined loss function was thus
\begin{equation}
\label{equ:comb_loss}
\begin{aligned}
    \mathcal{L}_{\mathrm{Combined}}(I_k, \hat{I}_k) 
    &= 
    \alpha \cdot \mathcal{L}_{\mathrm{DSC}}(I_k, \hat{I}_k) \\
    &\qquad + (1 - \alpha) \cdot \mathcal{L}_{\mathrm{CE}}(I_k, \hat{I}_k),
\end{aligned} 
\end{equation}
where $\mathcal{L}_{\mathrm{DSC}}$ denotes the soft \gls{dsc} loss~\cite{vnet-dice}, $\mathcal{L}_{\mathrm{CE}}$ denotes the \gls{ce} loss~\cite{cross-entropy}, and $\alpha \in [0, 1]$ is a parameter that determines the trade-off between the two losses. The value of $\alpha$ was found as part of the hyper-parameter search (see \secref{sec:opt} for details).

We will denote the ground truth delineations (that are annotated by a radiation oncologist) as \textit{delineations}, the network output predictions (by the single networks and the network using the \loss{}) as \textit{masks}, and the predictions by the teacher models in the \gls{ssl} (see \secref{sec:semi}) as \textit{pseudo masks}.

The function $\mathcal{L}_{\mathrm{DSC}}$ in \eqref{equ:comb_loss} denotes the soft \gls{dsc} loss, which is defined as~\cite{vu2019evaluation,vnet-dice,vu2020tunet,isensee2019nonew,vu2021multi}
\begin{equation} \label{eqn:dscloss}
    \mathcal{L}_{\mathrm{DSC}}(I, \hat{I}) = \frac{-2 \sum_l I_{l} \hat{I}_{l} + \epsilon }{\sum_l I_{l} + \sum_l \hat{I}_{l} + \epsilon},
\end{equation}
where the $I$ is the ground truth delineation,
the $\hat{I}$ is a predicted mask,
the sum is over the pixels in the delineations and masks,
and $\epsilon = 1 \cdot 10^{-5}$ is a small constant added to avoid division by zero and to make correctly predicted empty masks have a high \gls{dsc} score.

Further, the function $\mathcal{L}_{\mathrm{CE}}$ in \eqref{equ:comb_loss} denotes the \gls{ce} loss, which is defined as
\begin{equation} \label{eqn:celoss}
    \mathcal{L}_{\mathrm{CE}}(I, \hat{I}) = -\sum_l I_{l} \cdot \text{log} (\hat{I}_{l}).
\end{equation}

\begin{figure*}[!th]
    \centering
    \includegraphics[width=0.9\textwidth]{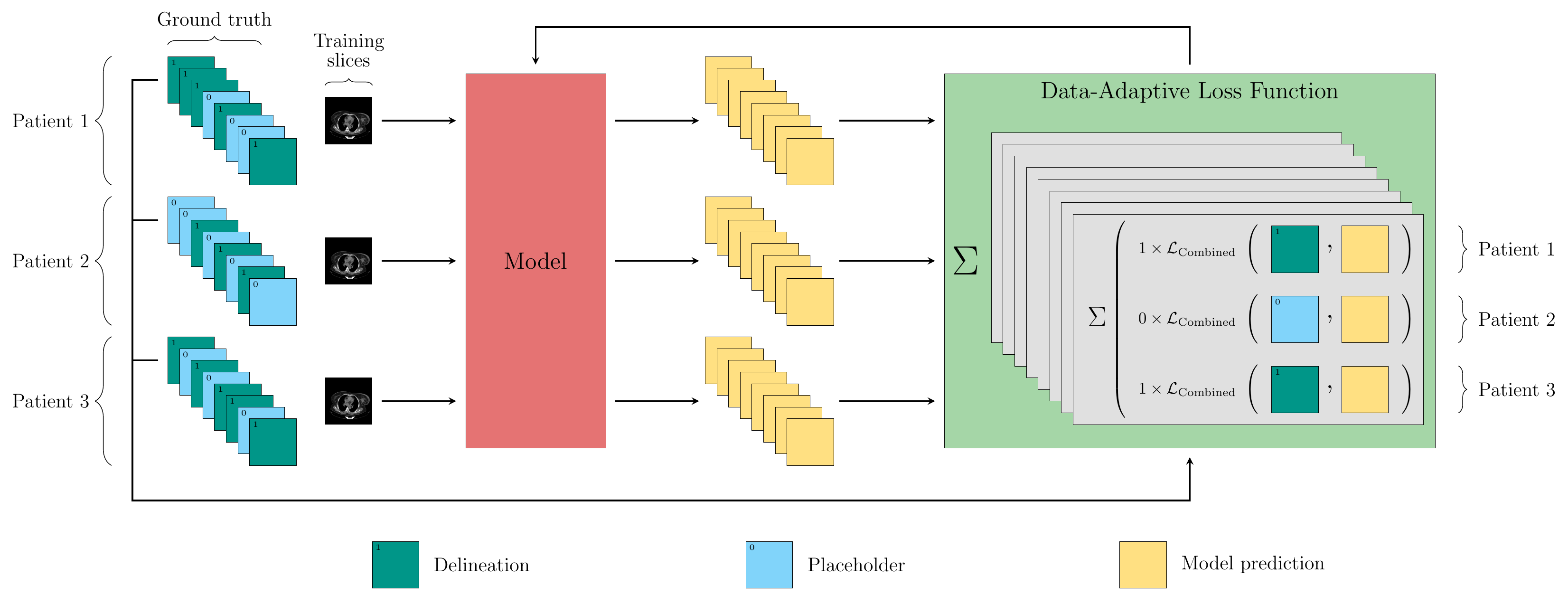}
    \caption{Illustration of the proposed method including a two-dimensional \gls{cnn} model together with the proposed data-adaptive loss function. The model is presented with ground truth delineations (dark green color) and placeholders (light blue color) during training. The placeholders are used here to indicate an image where the ground truth delineations were missing, and hence had weight zero (\ie, had $w_{i,j,k}=0$).
    }
    \label{fig:class_agnostic}
\end{figure*}

\begin{figure*}[!th]
    \centering
    \includegraphics[width=0.9\textwidth]{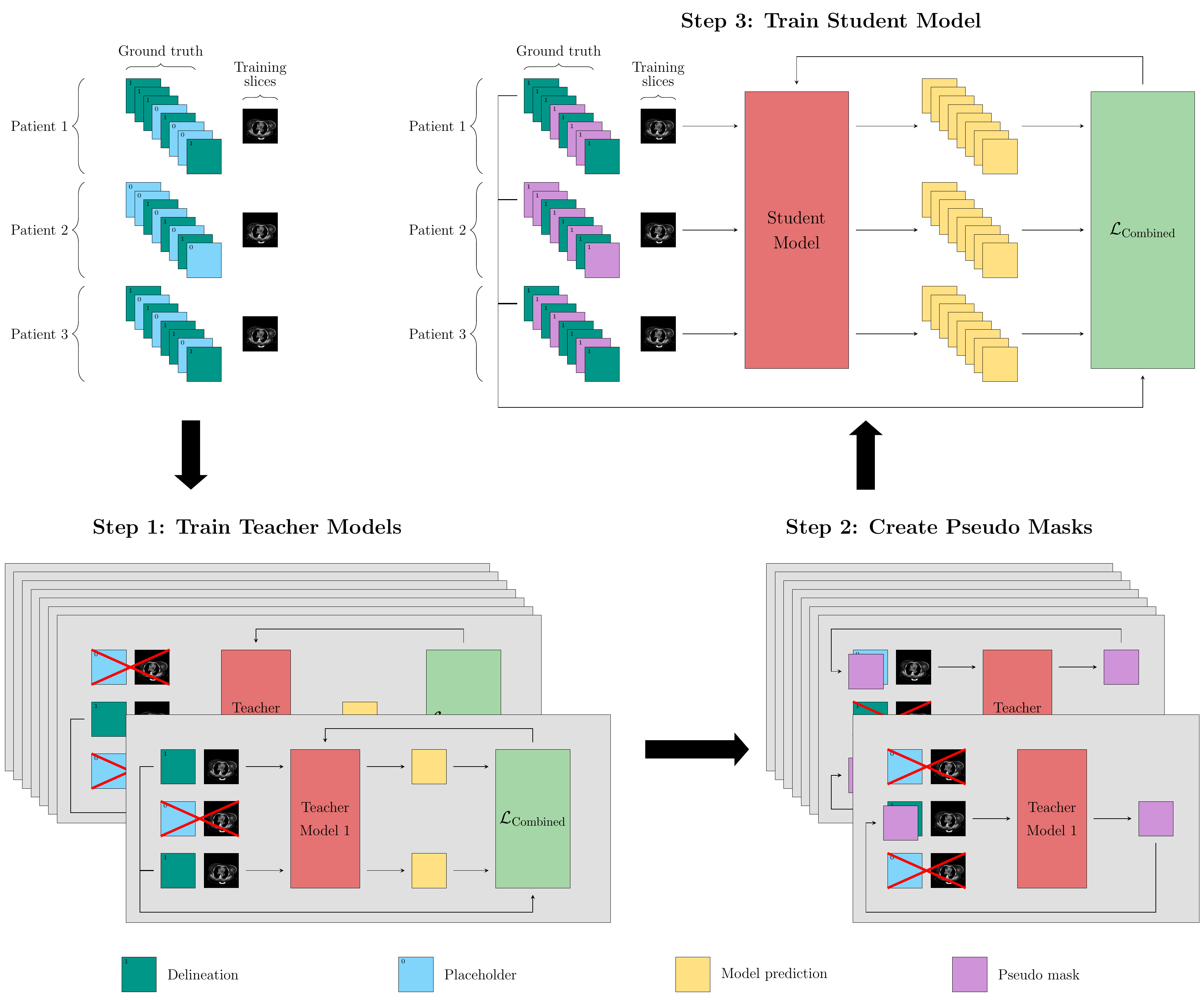}
    \caption{An illustration of the \gls{ssl} used in this work. The \gls{ssl} model has a \textit{teacher model} that is trained on the labeled data and then used to generate pseudo masks for the unlabelled data. It then also has a \textit{student model} that is trained on both the delineations and the pseudo masks.
    }
    \label{fig:semi}
\end{figure*}

Since the patients may have a different set of structures delineated, a \loss{} has to be able to adapt to the available masks. We propose to only incorporate the masks for which the ground truth delineations exist and normalize the loss appropriately in order to maintain the scale of the loss across different available ground truths. The \loss{} thus only accounts for the delineations in a given slice that actually exists. Hence, the model would only be updated with regards to the available delineations in any given step.

Let $w_{i,j,k}\in\{0,1\}$ denote whether or not structure $k=1,\ldots,8$ exists for slice $j$ in patient $i=1,\ldots,n$. This flag is thus used to indicate whether or not the $k$-th mask was included for a given patient slice, and the contribution of the masks to the loss (using \eqref{equ:comb_loss}) is an average over the available masks. Similarly, the contribution to the loss of a mini-batch is normalized by the number of a particular ground truth mask that were available for each patient slice in the mini-batch rather than the number of slices in the mini-batch. Hence, only masks that are available will induce a loss for a given patient slice. The proposed \loss{} for a mini-batch of patient slices is thus,
\begin{equation} \label{eqn:proposed}
    \begin{aligned}
        \mathcal{L}(I, \hat{I})
            & = \frac{
                \sum_{(i,j)\in\mathcal{I}}
                  \sum_{k=1}^8 w_{i,j,k} 
                  \cdot 
                  \mathcal{L}_{\mathrm{Combined}}(I_{i,j,k}, \hat{I}_{i,j,k})
            }{\sum_{(i,j)\in\mathcal{I}}\sum_{k=1}^8 w_{i,j,k}}, \\
    \end{aligned}
\end{equation}
where $\mathcal{I}$ is a set of patient and slice indices, $(i, j)$, in a mini-batch of delineations, $I$, and masks, $\hat{I}$, and where $I_{i,j}$ contains all delineations for the $j$-th slice of patient $i$, $\hat{I}_{i,j}$ contains all masks for the $j$-th slice of patient $i$, and $I_{i,j,k}$ and $\hat{I}_{i,j,k}$ are the $k$-th delineations and masks for slice $j$ of patient $i$, respectively. \figref{fig:class_agnostic} illustrates a 2D convolutional network together with the proposed loss function.

In order to tackle imbalanced data sets (see \secref{sec:dataset}), we also introduce two weighted \losspl{} that put appropriate emphasis on minority classes. First, the voxel-based weighted loss is defined as,
\begin{equation} \label{eqn:weightloss1}
    \begin{aligned}
        \mathcal{L_\mu}(I, \hat{I})
            & = \frac{
                \sum_{k=1}^8
                  \mu_k
                  \sum_{(i,j)\in\mathcal{I}}
                  w_{i,j,k}
                  \cdot 
                  \mathcal{L}_{\mathrm{Combined}}(I_{i,j,k}, \hat{I}_{i,j,k})
            }{\sum_{k=1}^8 \mu_k \sum_{(i,j)\in\mathcal{I}} w_{i,j,k}}, \\
    \end{aligned}
\end{equation}
where the $\mu_k\in(0,1]$ are weights for each structure, for $k=1,\ldots,8$, computed as exponentially weighted averages of the empirical frequencies of each class. We use bias correction after each update to the exponentially weighted averages.

Second, the slice-based weighted loss is defined as, 
\begin{equation} \label{eqn:weightloss2}
    \begin{aligned}
        \mathcal{L_\gamma}(I, \hat{I})
            & = \frac{
                \sum_{k=1}^8
                  \gamma_k
                  \sum_{(i,j)\in\mathcal{I}}
                  w_{i,j,k}
                  \cdot 
                  \mathcal{L}_{\mathrm{Combined}}(I_{i,j,k}, \hat{I}_{i,j,k})
            }{\sum_{k=1}^8 \gamma_k \sum_{(i,j)\in\mathcal{I}} w_{i,j,k}}, \\
    \end{aligned}
\end{equation}
where $\gamma_k\in(0,1]$ are weights for each structure, for $k=1,\ldots,8$, computed as exponentially weighted averages of the positive slices
for each class. Similar to the $\mu_k$, we also utilize bias correction after each update to the exponentially weighted averages.

\subsection{Data set} \label{sec:dataset}

The data used in this study were collected from $1\,614$ breast cancer patients at the University Hospital of Ume{\aa}, Ume{\aa}, Sweden. The data were collected between 2011 and the beginning of 2020 and contained \gls{ct} images and corresponding delineations of up to eight \glspl{oar}. The number of patient images for each target volume and \gls{oar} can be found in \tableref{table:data}. \tableref{table:data} also contains information about the number of patients in the training, validation, and test data sets.

\begin{table}[!th]
    \def\width{1.9 cm}
    \centering
    \begin{adjustbox}{max width=0.42\textwidth}
    \begin{tabular}{l c C{\width} C{\width} C{\width}}
        \toprule
        \multirow{3}{*}{Class}      && Train & Validation & Test \\ 
                                    && \# Patients & \# Patients & \# Patients \\ 
                                    && (\# Slices) & (\# Slices) & (\# Slices) \\
        \cmidrule{1-1}\cmidrule{3-5}
        Left breast                 && 343 (55\,151) & 149 (23\,774) & 31 (4\,963) \\ 
        Right breast                && 366 (57\,544) & 154 (24\,125) & 35 (5\,529)\\
        Left breast w. ax           && 180 (30\,460) &  75 (12\,753) & 13 (2\,175) \\
        Right breast w. ax          && 155 (26\,107) &  69 (11\,794) & 20 (3\,468) \\        
        Left lung                   && 552 (90\,705) & 240 (39\,193) &  46 (7\,500)\\ 
        Right lung                  && 541 (86\,929) & 233 (37\,565) & 57 (9\,396)\\
        Heart                       && 541 (86\,929) & 240 (39\,188) & 45 (7\,364) \\
        Spinal cord                 && 472 (69\,627) & 202 (30\,656) & 46 (6\,367) \\
        Spinal cord (*)             && 472 (81\,792) & 202 (34\,820) & 46 (7\,667) \\
        \cmidrule{1-1}\cmidrule{3-5}
        Total                       && \makecell{1\,059 \\ (171\,463)} 
        											 & \makecell{455 \\ (73\,745)} 
        											 				 & \makecell{100 \\ (16\,386)} \\
        \bottomrule
    \end{tabular}
    \end{adjustbox}
    \caption{The organs delineated in the data set, and their partition into training, validation, and test data sets. For the spinal cord, the first row indicates how many slices were delineated, and the second row (marked with an asterisk, $\ast$) denotes the total number of slices. It means that if all slices have delineations, the numbers in these two rows would thus be the same.
    }
    \label{table:data}
\end{table}

During imaging, a wire was sometimes placed around the patient's breast to guide the manual segmentation. This wire was removed during pre-processing by thresholding delineations and selecting the largest structure. The pixel values in the structure were replaced by the mean of the neighborhood pixel values. 

The \gls{ct} slices were down-sampled from $512 \times 512$ pixels to $256 \times 256$ pixels. Pixel values below $-1\,000$ Hounsfield units were set to $-1\,000$. The images were then rescaled by adding $1\,000$, and dividing by $3\,000$; \ie, the voxel values were (approximately) in the zero-one range.

\subsection{Baseline Models}

To evaluate the proposed \losspl{}, we first trained eight baseline models, one for each of the eight organs: left breast, right breast, left breast with lymph nodes, right breast with lymph nodes, left lung, right lung, heart, and spinal cord. Each baseline model had a modified U-Net architecture, based on Ronneberger \textit{et al.}~\cite{ronneberger2015u}, that was altered in the depth, number of filters, and additionally had spatial dropout layers. We used the combined loss function in \eqref{equ:comb_loss}. We further used Bayesian optimization to determine the hyper-parameters for each model (see \secref{sec:opt} and \tableref{table:parameter_search}).

\subsection{Semi-Supervised Model} \label{sec:semi}

The \gls{ssl} approach used here was inspired by that of Zhou \textit{et al.}~\cite{semi_1_Zhou}. The approach works as follows: A set of \textit{teacher models}, one for each structure (\ie, left breast, right breast, left breast with lymph nodes, right breast with lymph nodes, left lung, right lung, heart, and spinal cord), was trained on the available delineations for each \gls{oar}. In the present work, these teacher models were the same as the baseline models. The teacher models were then used to predict the missing delineations, giving a full set of labels containing both the available delineations and the predicted pseudo masks in the cases when the delineations were missing. Finally, a \textit{student model} was trained on all the data, \ie on both the ground truth delineations and on the pseudo masks. \figref{fig:semi} illustrates the \gls{ssl} approach that was used in this work.

\subsection{Hyper-Parameter Search} \label{sec:opt}

\begin{table*}[!th]
	\def\width{1.15 cm}
    \centering
    \begin{adjustbox}{max width=\textwidth}
	\begin{tabular}{l c c c C{\width} C{\width} C{\width} C{\width} C{\width} C{\width} C{\width} C{\width} C{\width} C{\width}}
       \toprule
        Hyper-parameter                      && Possible values 					&& Left breast		& Right breast		& Left breast w. ax 	& Right breast w. ax 	& Left lung & Right lung 	& Heart 	& Spinal cord 	& Semi supervised 	& Proposed 	\\
        \cmidrule{1-1} \cmidrule{3-3} \cmidrule{5-14} 
        Network depth                        && \{3, 4, 5\} 			&& 5				& 5					& 3						& 4						& 3			& 5				& 4			& 3			  	& 5				  	& 5			\\ 
        Base filters          				 && \{8, 16, 32, 64\} 		&& 32				& 32				& 8						& 32					& 16		& 32			& 16		& 32			& 32			  	& 32		\\
        Spatial dropout rate                 && [0.0, 0.6] 				&& 0.4541			& 0.1657			& 0.0420				& 0.7004				& 0.1740	& 0.7431		& 0.7074	& 0.6783		& 0.0696		  	& 0.3269	\\
        Optimizer                            && \{SGD, RMS, Adam\} 		&& Adam				& SGD				& RMS					& RMS					& Adam		& RMS			& RMS		& RMS			& RMS				& RMS		\\
        Trade-off factor, $\alpha$ 			 && [0.0, 1.0] 				&& 0.9934			& 0.9989			& 0.8871				& 0.8009				& 0.9843	& 0.9133		& 0.6269	& 0.9473		& 0.5264			& 0.3361	\\
        Mini-batch size                      && [1, 128] 				&& 13				& 27				& 126					& 19					& 33		& 31			& 42		& 8				& 28				& 28		\\
        Learning rate, $10^{\gamma}$         && [-6, -1] 				&& -3.8958			& -2.2390			& -3.9932				& -5.2986				& -3.2798	& -2.7654		& -1.4476	& -2.7624		& -3.1089  			& -5.5111	\\
        Number of epochs                     && [10, 120] 				&& 15				& 38				& 72					& 92					& 63		& 25			& 114		& 110			& 68				& 96		\\
       \bottomrule
	\end{tabular}
	\end{adjustbox}
	\caption{The search space for the hyper-parameter search. SGD denotes mini-batch stochastic gradient descent with momentum and RMS denotes the RMSprop optimizer. This table also contains the found values for evaluated methods.}
	\label{table:parameter_search}
\end{table*}

A hyper-parameters is defined as a model parameter that is not immediately updated by the optimization procedure. In \glspl{cnn}, these would be for instance parameters that determine the network architecture, such as, the number of layers in the network, the number of filters in each layer, or the optimization algorithm used to train the \glspl{cnn}. Current deep \glspl{cnn} have a large number of hyper-parameters. There are many procedures proposed to systematically find a good set of hyper-parameters. A commonly employed approach is Bayesian optimization, where expensive to compute black-box functions can be estimated and optimized~\cite{snoek2012practical}.
To find the hyper-parameters of the \gls{cnn} models used in the present work, we employed Bayesian optimization to determine the hyper-parameters of each of the previously described models. In particular, we used the \gls{tpe} proposed by Bergstra \textit{et al.}~\cite{10.5555/2986459.2986743}
through the \textit{hyperopt} library~\cite{hyperopt} to find a good set of hyper-parameters for each network. The network architecture used in the present work was a modified U-Net model~\cite{ronneberger2015u}. The hyper-parameter space over which we searched included:

\begin{itemize}
    \item Network depth: The depth of the modified U-Net.
    \item Base filters: The number of filters used in the first layer in the modified U-Net.
    \item Spatial dropout rate: The fraction of the input filters to drop in each step. To reduce overfitting, we added spatial dropout~\cite{spatialdropout} after each max-pooling or concatenation layer.
    \item Optimizer: The minimization algorithm used.
    \item Loss trade-off factor, $\alpha$: The parameter controlling the contribution of the \gls{dsc} and \gls{ce} losses to the combined loss (see \eqref{equ:comb_loss}).
    \item Mini-batch size: The number of slices included in each network update step.
    \item Learning rate: The (initial) step size used in the gradient descent-based optimization algorithms.
    \item Number of epochs: The number of times the entire training data set was presented to the model.
\end{itemize}

The parameter ranges or values used in the hyper-parameter search are listed in \tableref{table:parameter_search}. The hyper-parameter search was performed in $40$ iterations for each model.

\subsection{Incremental Learning}

We conducted experiments on incremental learning by first letting the proposed model \textit{learn} on $k-1$ classes, where $k$ was the total number of classes, and then extend the existing model's knowledge, \ie further trained the model on the left-out class. The purpose of incremental learning is to adapt a trained model to new data without forgetting its existing knowledge, and hence does not retrain the model from its initial configuration when new data arrives. In this work, we performed the incremental learning $k=8$ times, so that each structure was added once to a model trained on the $k-1$ other structures (see \tableref{tab:incremental_0123}). The set of hyper-parameters used in the incremental learning was the ones found in the hyper-parameter search for the model trained on all classes with the proposed loss function (see the ``Proposed'' column tabulated in \tableref{table:parameter_search}).


\subsection{Evaluation}
\label{subsec:eval}

This section describes the evaluation metrics that were used in this study to evaluate the segmentation performance. One of these metrics was the \gls{dsc} which is defined as
\begin{equation}\label{eq:dice}
    D(I, \hat{I}) = -\mathcal{L}_{\mathrm{DSC}}(I, \hat{I}),
\end{equation}
but with $\epsilon = 0$. The \gls{dsc} would thus be 1 if delineation and mask are the same (including the case that both delineation and mask are empty).

The segmentation performance was also evaluated using the \gls{hd95}, a common metric for evaluating segmentation performances. The \gls{hd} is defined as
\begin{equation}
    H(I, \hat{I}) = \max \big\{d(I, \hat{I}), d(\hat{I}, I)\big\},
\end{equation}
where 
\begin{equation}
    d(I, \hat{I}) = \max_{\vphantom{\hat{I}_{m}}I_{l} \in I} {\min_{\hat{I}_{m} \in \hat{I}} \|I_{l} - \hat{I}_{m}\|_2},
\end{equation}
in which $\|I_{l} - \hat{I}_{m}\|_2$ is the spatial Euclidean distance between pixels $I_{l}$ and $\hat{I}_{m}$ on the boundaries of the delineation $I$ and mask $\hat{I}$.

We also used the \gls{ravd} between the binary objects in the delineation and the mask. The \gls{ravd} is computed as the total volume difference of the delineation to the mask followed by the division by the total volume of the mask. The \gls{ravd} is defined as
\begin{equation*}
    R(I, \hat{I}) = 100 \cdot \frac{\sum_{x \in I} \delta_{x,1} - \sum_{x \in \hat{I}} \delta_{x,1}}{\sum_{x \in \hat{I}} \delta_{x,1}}
\end{equation*}
where $\delta_{x,1}$ denotes the Kronecker delta function, which takes the value 1 if $x=1$, and 0 otherwise.

The signed \gls{ravd} numbers are reported in \secref{sec:results}. A negative value is interpreted as under-segmentation and a positive value as over-segmentation.
To obtain a single score value, the absolute value is used. Note that a perfect value of zero can also be obtained for a non-perfect segmentation, as long as the volume of that segmentation is equal to the volume of the ground truth.

Finally, we computed the \gls{assd}. This metric is closely related to the \gls{hd95}, but instead of the 95$^\text{th}$ percentile, it computes the average \gls{asd} between the binary objects in the segmentation and the ground truth.

\subsection{Statistical tests} \label{sec:statistical}

To formally analyze the model performances, we used a Friedman test of equivalence between all evaluated methods on the evaluated metrics using the predictions on the test set. 
The Friedman test, when reporting significant differences, can be followed by a Nemenyi post-hoc test of pair-wise differences~\cite{Demsar2006}. 

\begin{table}[!th]
    \def\width{0.7 cm}
    \centering
    \begin{adjustbox}{max width=0.35\textwidth}
    \begin{tabular}{r C{\width} C{\width} C{\width} C{\width} C{\width}}
                                                        &
        \rotso{Single models}                           &
        \rotso{Semi-supervised}                         &
        \rotso{Proposed method \eqref{eqn:proposed}}    &
        \rotso{Voxel-based \eqref{eqn:weightloss1}}     &
        \rotso{Slice-based \eqref{eqn:weightloss2}}     \\
        \cmidrule{1-6}
        Single models                           & ~~~ & $+$ & $0$ & $0$ & $0$ \\
        Semi-supervised                         & $-$ & ~~~ & $-$ & $-$ & $-$ \\
        Proposed method \eqref{eqn:proposed}    & $0$ & $+$ & ~~~ & $0$ & $0$ \\
        Voxel-based \eqref{eqn:weightloss1}     & $0$ & $+$ & $0$ & ~~~ & $0$ \\
        Slice-based \eqref{eqn:weightloss2}     & $0$ & $+$ & $0$ & $0$ & ~~~ \\
        \bottomrule
    \end{tabular}
    \end{adjustbox}
    \caption{The results of the Nemenyi post-hoc test comparing all evaluated methods. A minus ($-$) means ranked significantly lower, a zero ($0$) means non-significant difference, and a plus ($+$) means ranked significantly higher, when comparing a method in the rows to a method in the columns.}
    \label{tab:result_nemenyi}
\end{table}

\section{Implementation Details and Training} \label{sec:details}

The models, the proposed loss, and the experiments overall were implemented using Keras 2.2.4\footnote{\url{https://keras.io}} with TensorFlow 1.12.0\footnote{\url{https://tensorflow.org}} as the backend. The models were trained on NVIDIA Tesla K80 and GeForce RTX 2080 Ti \glspl{gpu}. The training time for one hyper-parameter search was between 1--3 weeks.

The U-Net architecture has been a very successful architecture for medical imaging applications, and in particular, for semantic image segmentation~\cite{risk_organ_seg_zhikai, multi_organ_sulaiman, 3d_sparse_annotation}, and we, therefore, used a modified version of the U-Net in this project. Batch normalization was applied after every convolutional layer. 

To speed up the hyper-parameter search, we utilized asynchronous updates, allowing multiple trials to be evaluated in parallel. For a fair comparison, four \glspl{gpu} were simultaneously employed and the number of successful iterations was set to 40 for each hyper-parameter search. The objective functions used in the hyper-parameter search were different among models: the dice score was used for the eight baseline models, while the mean of the dice scores overall classes was used for the rest.

In the experiment with incremental learning, we used the models trained on $k-1$ \glspl{oar} as starting points. We then reused the weights for the $k-1$ already trained layers, and only initialized a new node for the additional $k$-th class in the last layer, meaning that the last layer in the new model now has $k$ nodes corresponding to $k$ structures. Finally, we trained the new model on the old data of the starting $k-1$ \glspl{oar} as well as the new data collected for the additional structure.

\section{Results} \label{sec:results}

The Friedman test reported a significant difference between the methods. The results from the Nemenyi post-hoc test are tabulated in \tableref{tab:result_nemenyi}. The test gives no significant differences between any methods, except for the \gls{ssl} which performed significantly worse than the other methods.

Except for in \tableref{tab:single}, we only report the results on the proposed model without weights (\ie, using \eqref{eqn:proposed}), since the differences were non-significant, per the Nemenyi test, between the method without weights and the voxel-based and slice-based methods (\ie, \eqref{eqn:weightloss1} and \eqref{eqn:weightloss2}, respectively).

\tableref{table:parameter_search} contains the eight hyper-parameters chosen for each model by the hyper-parameter search for the eight baseline models and the model with the proposed \loss{}. We see in \tableref{table:parameter_search} that each model ended up having a different set of hyper-parameters, as expected. Interestingly, RMSprop tended to be the most favored optimization algorithm.

\begin{table}[!th]
  	\def\width{1.9 cm}
    \centering
    \begin{adjustbox}{max width=0.49\textwidth}
    \begin{tabular}{l c C{\width} C{\width} C{\width} C{\width}}
        \toprule
        Class                   && DSC                 & HD95                & RAVD                & ASSD  \\
        
        \cmidrule{1-1} \cmidrule{3-6}
        \multicolumn{6}{l}{Single models} \\       
        \cmidrule{1-1} \cmidrule{3-6}
		Left breast             && 0.88 (0.004)        & 3.66 (0.191)        & 0.09 (0.068)        & 1.63 (0.126)       \\
		Right breast            && 0.92 (0.003)        & 2.09 (0.044)        & 0.15 (0.091)        & 0.64 (0.014)       \\
		Left breast w.   ax     && 0.83 (0.006)        & 4.62 (0.164)        & 0.17 (0.125)        & 1.45 (0.090)       \\
		Right breast w.   ax    && 0.84 (0.004)        & 4.13 (0.080)        & 0.03 (0.032)        & 1.30 (0.030)       \\
		Left lung               && 0.97 (0.001)        & 1.43 (0.061)        & 0.01 (0.006)        & 0.29 (0.022)       \\
		Right lung              && 0.97 (0.001)        & 1.51 (0.055)        & 0.00 (0.006)        & 0.23 (0.010)       \\
		Heart                   && 0.94 (0.002)        & 1.28 (0.046)        & 0.19 (0.089)        & 0.44 (0.017)       \\
		Spinal cord             && 0.76 (0.004)        & 1.05 (0.019)        & -0.15 (0.012)       & 0.36 (0.007)       \\
		
        \cmidrule{1-1} \cmidrule{3-6}
        \multicolumn{6}{l}{Semi-supervised} \\       
        \cmidrule{1-1} \cmidrule{3-6}        
        Left breast             && 0.78 (0.005)        & 5.07 (0.203)        & 2.70 (0.535)        & 1.93 (0.121)       \\
        Right breast            && 0.82 (0.004)        & 3.76 (0.101)        & 1.27 (0.260)        & 1.12 (0.043)       \\
        Left breast with ax     && 0.64 (0.009)        & 7.90 (0.285)        & 7.71 (1.943)        & 3.09 (0.139)       \\
        Right breast with ax    && 0.60 (0.007)        & 7.76 (0.247)        & 6.73 (1.192)        & 3.17 (0.120)       \\
        Left lung               && 0.88 (0.003)        & 3.77 (0.135)        & 5.98 (0.898)        & 1.15 (0.061)       \\
        Right lung              && 0.90 (0.003)        & 2.73 (0.089)        & 3.02 (0.733)        & 0.60 (0.032)       \\
        Heart                   && 0.89 (0.003)        & 1.80 (0.060)        & 0.55 (0.284)        & 0.60 (0.023)       \\
        Spinal cord             && 0.70 (0.004)        & 1.81 (0.039)        & -0.18 (0.012)       & 0.59 (0.007)       \\    
        
        \cmidrule{1-1} \cmidrule{3-6}
        \multicolumn{6}{l}{Proposed method (see \eqref{eqn:proposed})} \\       
        \cmidrule{1-1} \cmidrule{3-6}        
		Left breast             && 0.88 (0.004)        & 3.65 (0.187)        & -0.01 (0.015)       & 1.69 (0.130)       \\
		Right breast            && 0.92 (0.003)        & 1.98 (0.043)        & -0.00 (0.005)       & 0.65 (0.015)       \\
		Left breast w.   ax     && 0.85 (0.005)        & 4.12 (0.110)        & 0.14 (0.098)        & 1.32 (0.037)       \\
		Right breast w.   ax    && 0.86 (0.004)        & 4.38 (0.128)        & 0.03 (0.009)        & 1.21 (0.028)       \\
		Left lung               && 0.97 (0.001)        & 1.60 (0.066)        & 0.02 (0.016)        & 0.28 (0.021)       \\
		Right lung              && 0.97 (0.001)        & 1.86 (0.069)        & -0.00 (0.005)       & 0.21 (0.005)       \\
		Heart                   && 0.92 (0.003)        & 1.38 (0.046)        & -0.04 (0.006)       & 0.47 (0.015)       \\
		Spinal cord             && 0.75 (0.004)        & 1.06 (0.010)        & -0.14 (0.018)       & 0.38 (0.004)       \\
		
        \cmidrule{1-1} \cmidrule{3-6}
        \multicolumn{6}{l}{Proposed voxel-based method (see \eqref{eqn:weightloss1})} \\      
        \cmidrule{1-1} \cmidrule{3-6}        		
        Left breast             && 0.87 (0.004)        & 3.82 (0.191)        & 0.10 (0.148)        & 1.71 (0.131)       \\
        Right breast            && 0.91 (0.003)        & 2.16 (0.049)        & -0.00 (0.010)       & 0.65 (0.016)       \\
        Left breast with ax     && 0.86 (0.005)        & 4.09 (0.103)        & 0.01 (0.036)        & 1.25 (0.033)       \\
        Right breast with ax    && 0.86 (0.004)        & 3.82 (0.079)        & 0.00 (0.045)        & 1.14 (0.026)       \\
        Left lung               && 0.97 (0.001)        & 2.00 (0.071)        & 0.00 (0.006)        & 0.29 (0.019)       \\
        Right lung              && 0.97 (0.001)        & 1.92 (0.059)        & -0.00 (0.004)       & 0.24 (0.005)       \\
        Heart                   && 0.93 (0.002)        & 1.40 (0.046)        & -0.00 (0.014)       & 0.48 (0.015)       \\
        Spinal cord             && 0.75 (0.004)        & 1.06 (0.007)        & -0.10 (0.007)       & 0.41 (0.004)       \\	
		
        \cmidrule{1-1} \cmidrule{3-6}
        \multicolumn{6}{l}{Proposed slice-based method (see \eqref{eqn:weightloss2})} \\       
        \cmidrule{1-1} \cmidrule{3-6}        		
        Left breast             && 0.88 (0.004)        & 3.72 (0.189)        & 0.09 (0.049)        & 1.77 (0.133)       \\
        Right breast            && 0.92 (0.003)        & 2.17 (0.051)        & 0.10 (0.016)        & 0.76 (0.020)       \\
        Left breast with ax     && 0.86 (0.005)        & 3.96 (0.104)        & 0.14 (0.068)        & 1.25 (0.036)       \\
        Right breast with ax    && 0.87 (0.004)        & 3.56 (0.073)        & 0.07 (0.039)        & 1.16 (0.025)       \\
        Left lung               && 0.93 (0.003)        & 1.79 (0.070)        & -0.02 (0.019)       & 0.28 (0.019)       \\
        Right lung              && 0.97 (0.001)        & 1.62 (0.053)        & -0.01 (0.004)       & 0.25 (0.007)       \\
        Heart                   && 0.93 (0.002)        & 1.33 (0.045)        & 0.04 (0.020)        & 0.47 (0.017)       \\
        Spinal cord             && 0.75 (0.004)        & 1.06 (0.010)        & -0.17 (0.005)       & 0.39 (0.003)       \\       
        
        \bottomrule
    \end{tabular}
	\end{adjustbox}
    \caption{Mean \gls{dsc} (higher is better), \gls{hd95} (lower is better), \gls{ravd} (lower is better), and \gls{assd} (lower is better) and their \glspl{se} (in parentheses) computed on the test set from eight single models, \gls{ssl} model and the full model using the proposed loss functions.}
    \label{tab:single}
\end{table}

\tableref{tab:single} provides the mean \gls{dsc}, \gls{hd95}, \gls{ravd}, and \gls{assd} and their \glspl{se} (in parentheses) computed on the test set. The metrics were computed on (i) the eight single models, that were trained on the eight different classes independently, (ii) the \gls{ssl}, and (iii) the model using the proposed loss function.

\tableref{tab:incremental_0123} shows the mean \gls{dsc}, \gls{hd95}, \gls{ravd} and \gls{assd} and their \glspl{se} (in parentheses) of $k-1$ structures (before) and $k$ structures (after) using incremental learning computed on the test set of eight evaluated structures.

\begin{figure*}[!!htbp]
    \def\width{0.32\textwidth}
     \centering
     \begin{subfigure}[b]{\width}
         \centering
         \includegraphics[width=\textwidth]{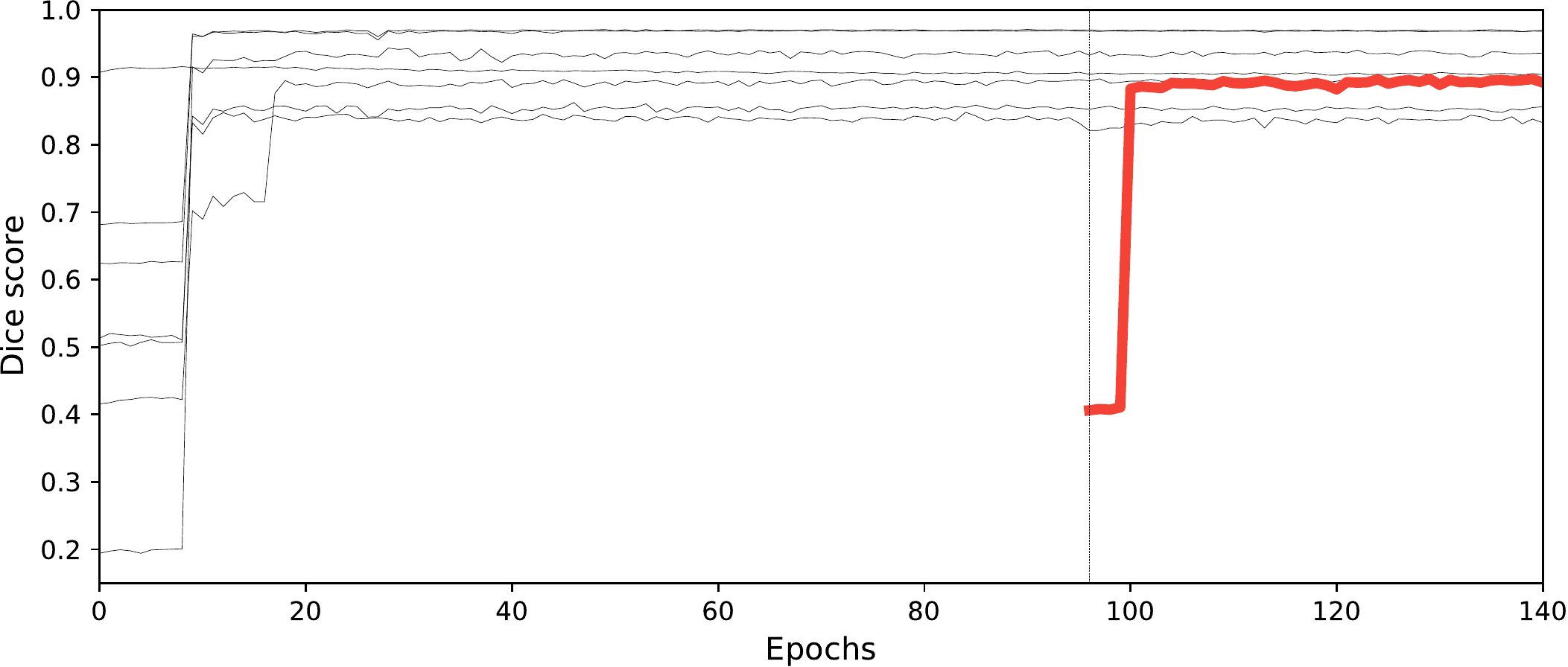}
         \caption{Left breast}
     \end{subfigure}
     \begin{subfigure}[b]{\width}
         \centering
         \includegraphics[width=\textwidth]{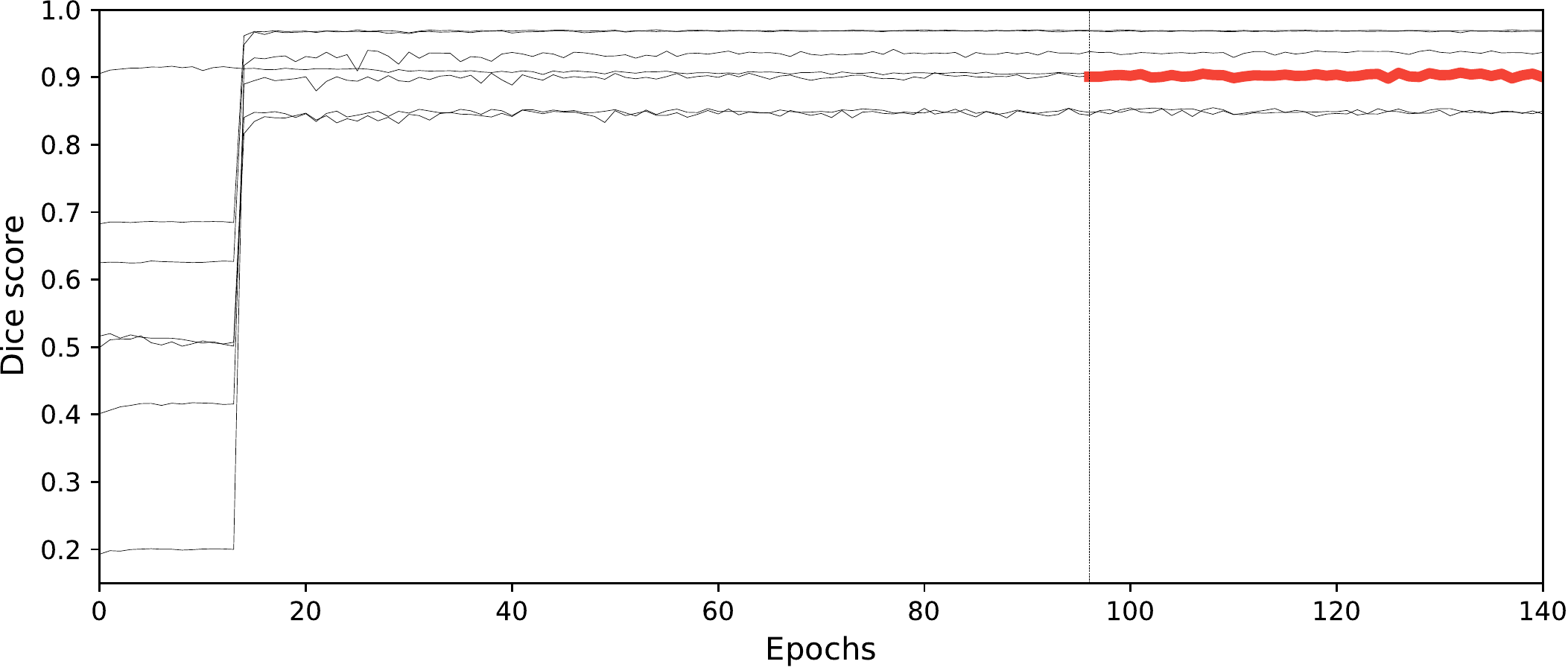}
         \caption{Right breast}
     \end{subfigure}
     \begin{subfigure}[b]{\width}
         \centering
         \includegraphics[width=\textwidth]{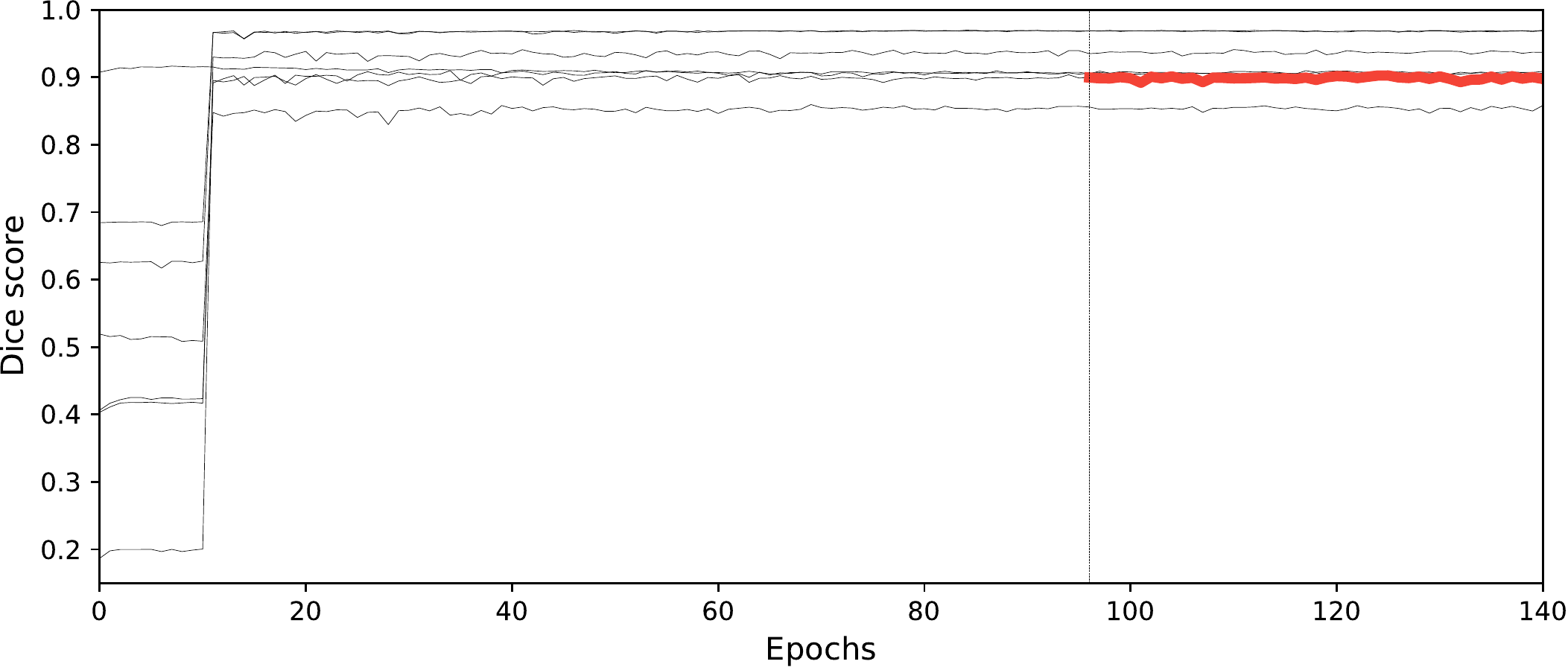}
         \caption{Left breast w. ax}
     \end{subfigure}
     \\[2mm]
     \begin{subfigure}[b]{\width}
         \centering
         \includegraphics[width=\textwidth]{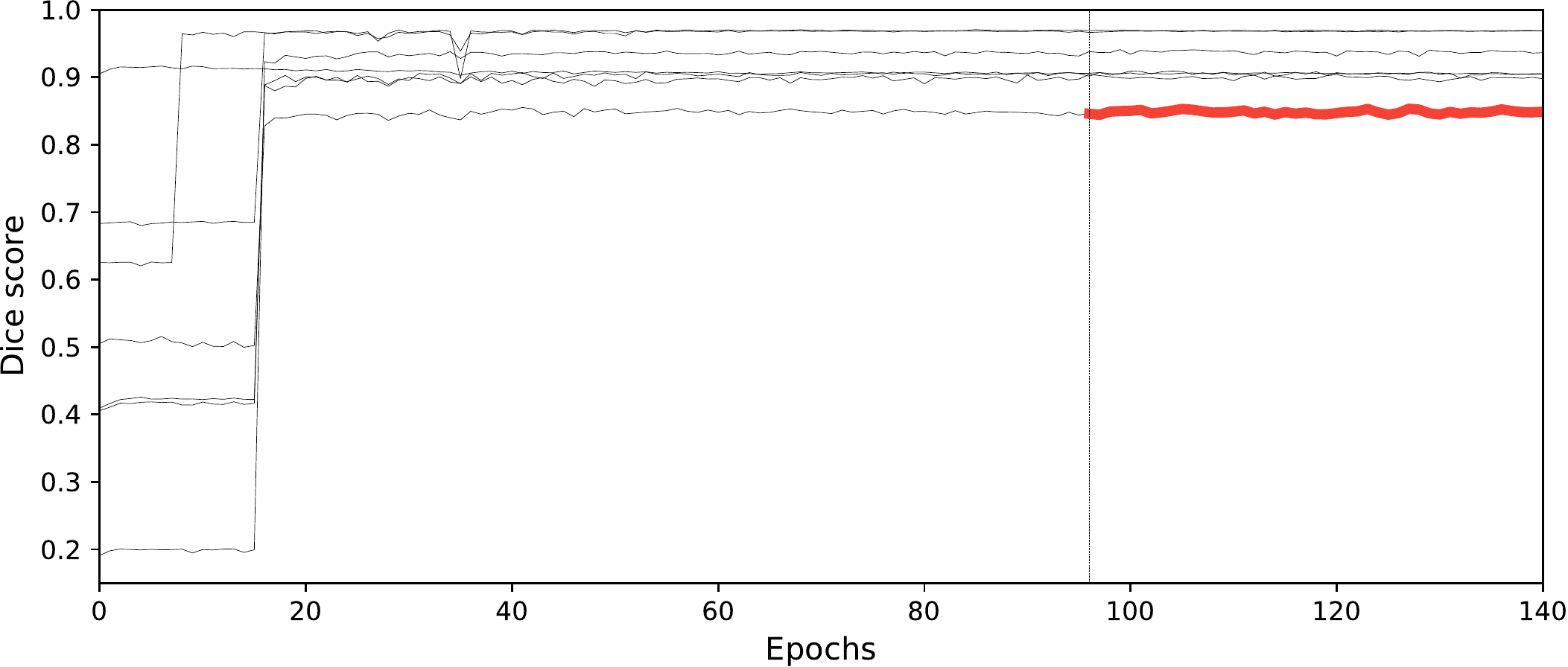}
         \caption{Right breast w. ax}
     \end{subfigure}        
     \begin{subfigure}[b]{\width}
         \centering
         \includegraphics[width=\textwidth]{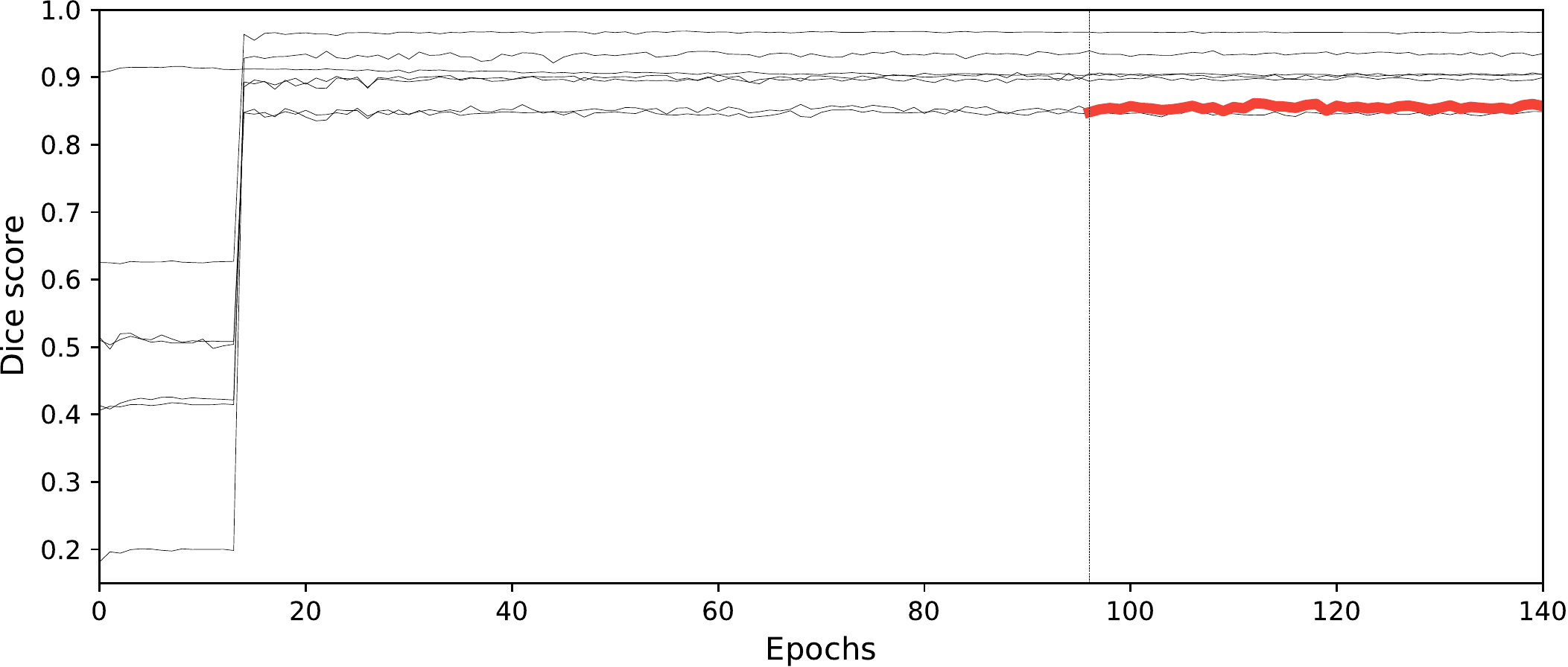}
         \caption{Left lung}
     \end{subfigure}
     \begin{subfigure}[b]{\width}
         \centering
         \includegraphics[width=\textwidth]{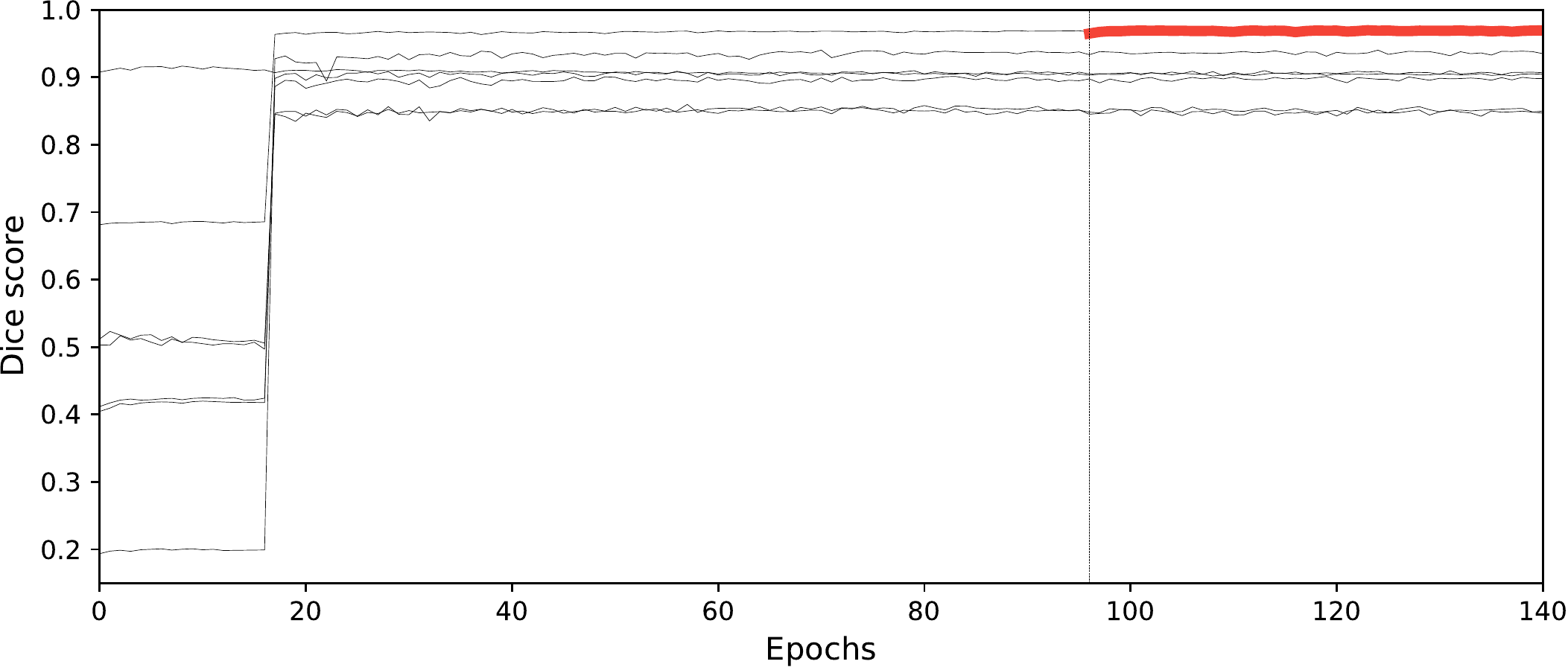}
         \caption{Right lung}
     \end{subfigure} 
     \\[2mm]
     \begin{subfigure}[b]{\width}
         \centering
         \includegraphics[width=\textwidth]{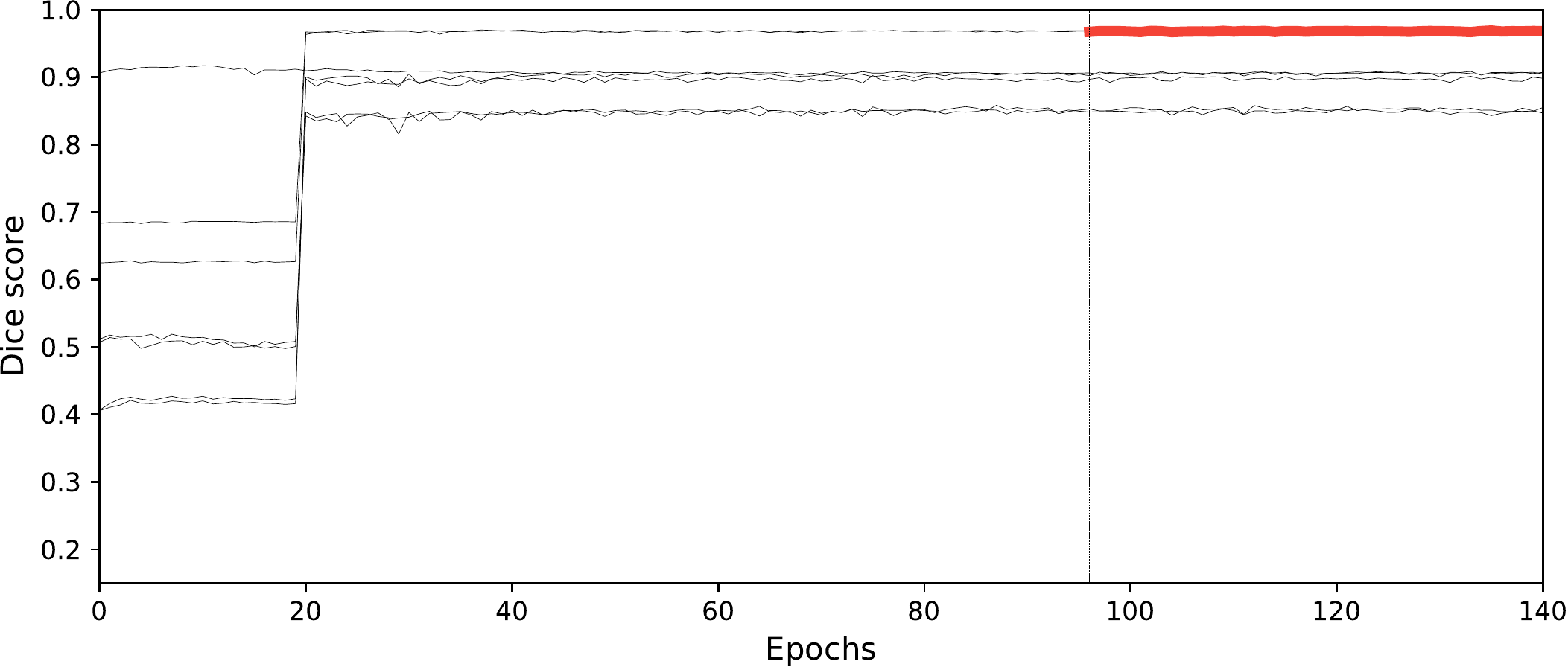}
         \caption{Heart}
     \end{subfigure}
     \begin{subfigure}[b]{\width}
         \centering
         \includegraphics[width=\textwidth]{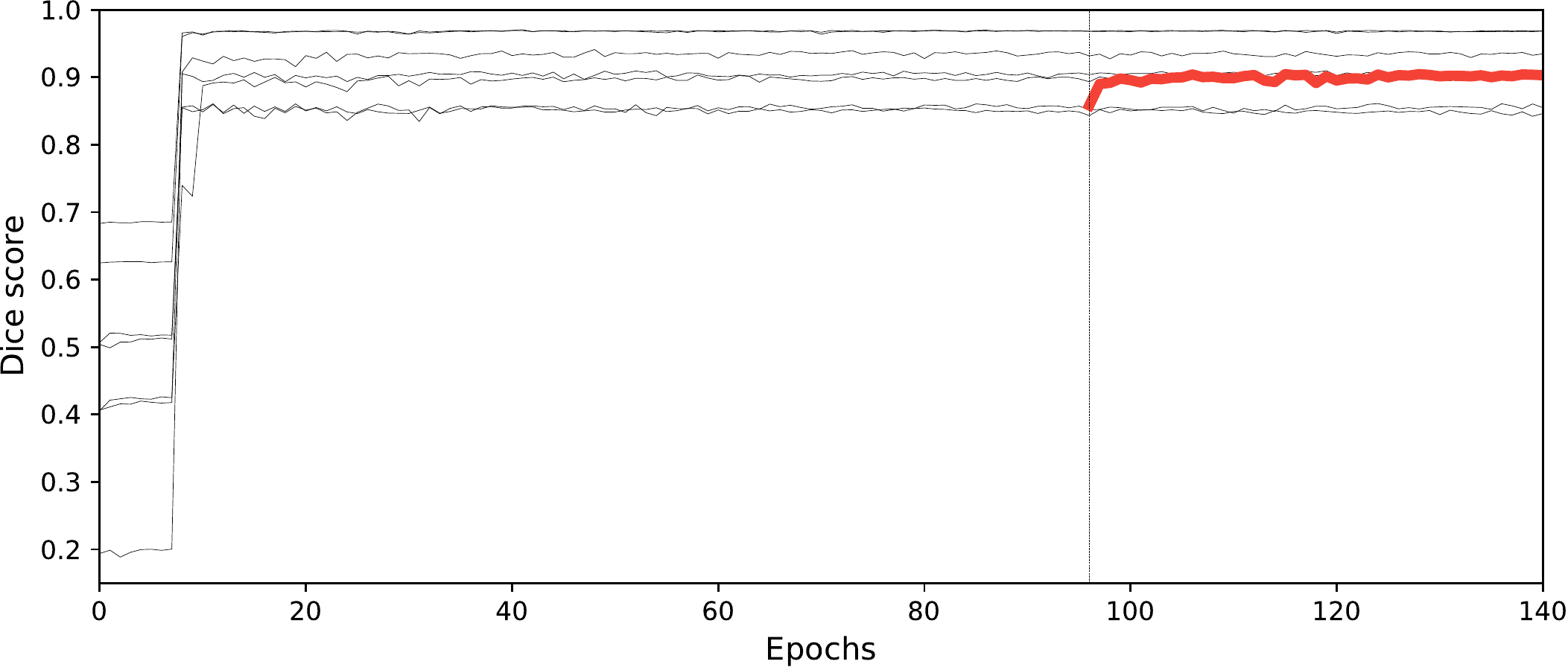}
         \caption{Spinal cord}
     \end{subfigure}      
        
     \caption{Illustration of the learning curve showing \gls{dsc} \textit{vs.}~epoch. In the incremental learning, $k-1$ \glspl{oar} were trained from the beginning, while an additional \gls{oar} was added starting at epoch 96 (vertical line). The initial structures are shown in black color, while the added \gls{oar} is displayed with a thick red line.} 
     \label{fig:learn_curve}
\end{figure*}

\figref{fig:learn_curve} presents the learning curves, showing how the \gls{dsc} changes as a function of the epoch number during training. Illustrated are the learning curves for $k-1$ \glspl{oar} from the beginning of training, and the \gls{oar} added at epoch 96 and its development until the end of the training. The initial \glspl{oar} are illustrated in black color, while the added structure is illustrated with a thicker red line.

\begin{figure*}[!!htbp]
    \centering
    \includegraphics[width=0.95\textwidth]{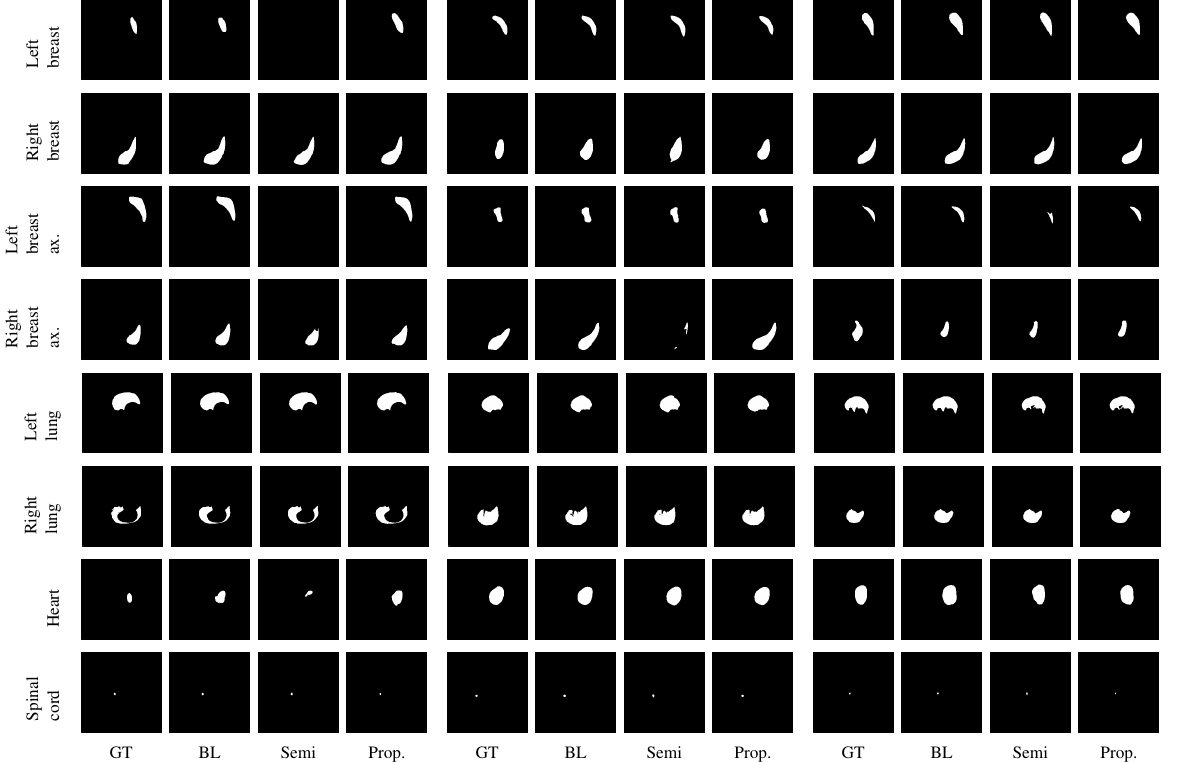}
    \caption{Qualitative results of the proposed method on eight classes: left breast, right breast, left breast with lymph nodes, right breast with lymph nodes, left lung, right lung, heart, and spinal cord. From left to right: delineation (GT), baseline model (BL), \gls{ssl} model (Semi), and the proposed method (Prop.).}
    \label{fig:qualitative}
\end{figure*}

\figref{fig:qualitative} illustrates the qualitative results of the baseline model, the \gls{ssl} model, and the proposed model on eight classes. Note that the image samples in \figref{fig:qualitative} were randomly selected, and are mainly for illustrative purposes. However, there are still a few observations that can be made, that are related to the quantitative results in \tableref{tab:single}, and are further discussed in \secref{sec:discussion}.

\begin{table*}[!htbp]
  	\def\width{1.9 cm}
    \centering
    \begin{adjustbox}{max width=0.95\textwidth}
    \begin{tabular}{l c C{\width} C{\width} C{\width} C{\width} c C{\width} C{\width} C{\width} C{\width}}
        \toprule
        \multirow{2}{*}{Class}  && \multicolumn{4}{c}{Before}                                                        && \multicolumn{4}{c}{After}                                                   \\
                                && DSC                & HD95               & RAVD               & ASSD                && DSC                 & HD95                & RAVD                & ASSD       \\
        \cmidrule{1-1} \cmidrule{3-6} \cmidrule{8-11}
		Left breast             &&                 	  &                    &                    &                    && 0.87 (0.004)        & 4.21 (0.209)        & -0.01 (0.013)       & 1.68 (0.125)       \\
		Right breast            && 0.93 (0.002)       & 2.18 (0.052)       & 0.06 (0.017)       & 0.65 (0.016)       && 0.90 (0.003)        & 1.83 (0.044)        & 0.02 (0.007)        & 0.71 (0.019)       \\
		Left breast w.   ax     && 0.86 (0.005)       & 4.36 (0.117)       & 0.03 (0.010)       & 1.34 (0.040)       && 0.86 (0.005)        & 4.32 (0.116)        & 0.17 (0.102)        & 1.36 (0.041)       \\
		Right breast w.   ax    && 0.89 (0.003)       & 3.87 (0.088)       & 0.07 (0.019)       & 1.24 (0.029)       && 0.84 (0.004)        & 4.17 (0.100)        & 0.06 (0.012)        & 1.31 (0.031)       \\
		Left lung               && 0.97 (0.001)       & 1.53 (0.063)       & 0.00 (0.005)       & 0.29 (0.021)       && 0.97 (0.001)        & 1.58 (0.065)        & 0.03 (0.013)        & 0.29 (0.021)       \\
		Right lung              && 0.97 (0.001)       & 1.57 (0.054)       & -0.01 (0.002)      & 0.22 (0.006)       && 0.97 (0.001)        & 1.65 (0.055)        & 0.00 (0.003)        & 0.23 (0.008)       \\
		Heart                   && 0.93 (0.003)       & 1.37 (0.045)       & -0.01 (0.004)      & 0.47 (0.014)       && 0.93 (0.002)        & 1.30 (0.041)        & 0.00 (0.011)        & 0.49 (0.015)       \\
		Spinal cord             && 0.75 (0.004)       & 1.07 (0.008)       & -0.11 (0.020)      & 0.39 (0.005)       && 0.73 (0.004)        & 1.21 (0.011)        & -0.14 (0.010)       & 0.41 (0.005)       \\
		\cmidrule{1-1} \cmidrule{3-6} \cmidrule{8-11}
		Left breast             && 0.88 (0.004)       & 3.79 (0.192)       & 0.03 (0.039)       & 1.70 (0.127)       && 0.87 (0.004)        & 3.79 (0.188)        & -0.01 (0.014)       & 1.72 (0.129)       \\
		Right breast            &&                    &                    &                    &                    && 0.91 (0.003)        & 2.05 (0.049)        & 0.02 (0.006)        & 0.72 (0.019)       \\
		Left breast w.   ax     && 0.88 (0.004)       & 4.10 (0.113)       & 0.05 (0.012)       & 1.33 (0.039)       && 0.86 (0.005)        & 3.64 (0.097)        & 0.16 (0.095)        & 1.39 (0.043)       \\
		Right breast w.   ax    && 0.88 (0.003)       & 4.07 (0.095)       & 0.11 (0.042)       & 1.28 (0.030)       && 0.84 (0.004)        & 3.55 (0.086)        & 0.06 (0.011)        & 1.33 (0.032)       \\
		Left lung               && 0.98 (0.001)       & 1.63 (0.066)       & 0.01 (0.008)       & 0.28 (0.021)       && 0.98 (0.001)        & 1.64 (0.068)        & 0.02 (0.010)        & 0.30 (0.022)       \\
		Right lung              && 0.97 (0.001)       & 1.88 (0.072)       & -0.01 (0.004)      & 0.21 (0.006)       && 0.97 (0.001)        & 1.66 (0.055)        & 0.00 (0.003)        & 0.24 (0.011)       \\
		Heart                   && 0.92 (0.003)       & 1.39 (0.046)       & 0.02 (0.024)       & 0.46 (0.014)       && 0.93 (0.002)        & 1.29 (0.041)        & 0.00 (0.009)        & 0.49 (0.015)       \\
		Spinal cord             && 0.75 (0.004)       & 1.07 (0.007)       & -0.15 (0.008)      & 0.40 (0.005)       && 0.72 (0.004)        & 1.01 (0.008)        & -0.14 (0.015)       & 0.40 (0.005)       \\
		\cmidrule{1-1} \cmidrule{3-6} \cmidrule{8-11}
		Left breast             && 0.88 (0.004)       & 3.45 (0.181)       & 0.01 (0.016)       & 1.66 (0.124)       && 0.87 (0.004)        & 3.95 (0.196)        & -0.01 (0.012)       & 1.71 (0.128)       \\
		Right breast            && 0.92 (0.003)       & 2.06 (0.044)       & 0.01 (0.005)       & 0.65 (0.015)       && 0.91 (0.003)        & 2.47 (0.059)        & 0.02 (0.008)        & 0.69 (0.017)       \\
		Left breast w.   ax     &&                    &                    &                    &                    && 0.85 (0.005)        & 4.16 (0.111)        & 0.18 (0.106)        & 1.31 (0.039)       \\
		Right breast w.   ax    && 0.88 (0.003)       & 3.71 (0.082)       & 0.04 (0.015)       & 1.18 (0.027)       && 0.87 (0.004)        & 3.52 (0.087)        & 0.05 (0.010)        & 1.27 (0.030)       \\
		Left lung               && 0.98 (0.001)       & 1.53 (0.063)       & 0.02 (0.010)       & 0.28 (0.020)       && 0.96 (0.001)        & 1.86 (0.076)        & 0.03 (0.016)        & 0.29 (0.021)       \\
		Right lung              && 0.97 (0.001)       & 1.85 (0.072)       & -0.00 (0.005)      & 0.21 (0.006)       && 0.96 (0.001)        & 1.41 (0.048)        & 0.00 (0.004)        & 0.22 (0.008)       \\
		Heart                   && 0.92 (0.003)       & 1.44 (0.048)       & 0.03 (0.028)       & 0.47 (0.014)       && 0.93 (0.002)        & 1.53 (0.049)        & 0.00 (0.011)        & 0.50 (0.016)       \\
		Spinal cord             && 0.75 (0.004)       & 1.10 (0.015)       & -0.18 (0.005)      & 0.39 (0.005)       && 0.76 (0.004)        & 1.07 (0.011)        & -0.15 (0.010)       & 0.39 (0.005)       \\
		\cmidrule{1-1} \cmidrule{3-6} \cmidrule{8-11}
		Left breast             && 0.88 (0.004)       & 3.55 (0.184)       & -0.00 (0.015)      & 1.74 (0.130)       && 0.87 (0.004)        & 4.26 (0.212)        & -0.01 (0.013)       & 1.72 (0.129)       \\
		Right breast            && 0.91 (0.003)       & 2.38 (0.062)       & 0.09 (0.034)       & 0.68 (0.016)       && 0.90 (0.003)        & 2.46 (0.059)        & 0.02 (0.007)        & 0.72 (0.019)       \\
		Left breast w.   ax     && 0.85 (0.005)       & 4.35 (0.131)       & 0.57 (0.358)       & 1.30 (0.038)       && 0.86 (0.005)        & 4.41 (0.119)        & 0.19 (0.107)        & 1.36 (0.043)       \\
		Right breast w.   ax    &&                    &                    &                    &                    && 0.84 (0.004)        & 4.65 (0.114)        & 0.06 (0.011)        & 1.30 (0.031)       \\
		Left lung               && 0.98 (0.001)       & 1.50 (0.061)       & 0.00 (0.004)       & 0.28 (0.021)       && 0.94 (0.001)        & 1.79 (0.073)        & 0.01 (0.006)        & 0.28 (0.021)       \\
		Right lung              && 0.97 (0.001)       & 1.63 (0.057)       & -0.02 (0.002)      & 0.22 (0.007)       && 0.96 (0.001)        & 1.91 (0.065)        & 0.00 (0.004)        & 0.24 (0.011)       \\
		Heart                   && 0.92 (0.003)       & 1.34 (0.045)       & 0.00 (0.015)       & 0.47 (0.014)       && 0.93 (0.002)        & 1.19 (0.038)        & 0.00 (0.011)        & 0.50 (0.016)       \\
		Spinal cord             && 0.75 (0.004)       & 1.19 (0.057)       & -0.11 (0.024)      & 0.40 (0.005)       && 0.74 (0.004)        & 0.95 (0.008)        & -0.15 (0.011)       & 0.40 (0.005)       \\	
        \cmidrule{1-1} \cmidrule{3-6} \cmidrule{8-11}
		Left breast             && 0.88 (0.004)       & 3.83 (0.187)       & -0.01 (0.011)      & 1.75 (0.131)       && 0.88 (0.004)        & 3.45 (0.172)        & -0.01 (0.016)       & 1.74 (0.130)       \\
		Right breast            && 0.91 (0.003)       & 2.46 (0.085)       & 0.25 (0.059)       & 0.67 (0.017)       && 0.90 (0.003)        & 2.11 (0.050)        & 0.02 (0.009)        & 0.71 (0.019)       \\
		Left breast w.   ax     && 0.85 (0.005)       & 4.56 (0.126)       & 0.06 (0.017)       & 1.31 (0.038)       && 0.83 (0.005)        & 4.82 (0.128)        & 0.14 (0.090)        & 1.35 (0.040)       \\
		Right breast w.   ax    && 0.87 (0.004)       & 3.70 (0.077)       & 0.06 (0.021)       & 1.23 (0.029)       && 0.87 (0.004)        & 3.60 (0.086)        & 0.05 (0.010)        & 1.32 (0.032)       \\
		Left lung               &&                    &                    &                    &                    && 0.95 (0.001)        & 1.57 (0.065)        & 0.02 (0.006)        & 0.29 (0.021)       \\
		Right lung              && 0.97 (0.001)       & 1.59 (0.054)       & 0.00 (0.005)       & 0.22 (0.007)       && 0.95 (0.001)        & 1.77 (0.061)        & 0.00 (0.003)        & 0.24 (0.012)       \\
		Heart                   && 0.92 (0.003)       & 1.35 (0.045)       & -0.02 (0.005)      & 0.46 (0.014)       && 0.92 (0.002)        & 1.59 (0.051)        & -0.00 (0.009)       & 0.49 (0.015)       \\
		Spinal cord             && 0.75 (0.004)       & 1.10 (0.014)       & -0.15 (0.008)      & 0.39 (0.005)       && 0.76 (0.004)        & 1.12 (0.011)        & -0.14 (0.013)       & 0.41 (0.005)       \\
		\cmidrule{1-1} \cmidrule{3-6} \cmidrule{8-11}
		Left breast             && 0.89 (0.004)       & 3.50 (0.182)       & 0.00 (0.016)       & 1.75 (0.131)       && 0.89 (0.004)        & 4.42 (0.220)        & -0.01 (0.015)       & 1.75 (0.131)       \\
		Right breast            && 0.91 (0.003)       & 2.02 (0.047)       & -0.01 (0.004)      & 0.67 (0.016)       && 0.91 (0.002)        & 1.73 (0.041)        & 0.02 (0.008)        & 0.71 (0.019)       \\
		Left breast w.   ax     && 0.86 (0.005)       & 4.19 (0.129)       & 0.10 (0.045)       & 1.34 (0.039)       && 0.85 (0.005)        & 3.97 (0.106)        & 0.17 (0.095)        & 1.42 (0.045)       \\
		Right breast w.   ax    && 0.86 (0.004)       & 3.83 (0.082)       & 0.01 (0.009)       & 1.25 (0.029)       && 0.86 (0.004)        & 4.04 (0.099)        & 0.05 (0.010)        & 1.26 (0.030)       \\
		Left lung               && 0.97 (0.001)       & 1.79 (0.073)       & -0.00 (0.004)      & 0.28 (0.021)       && 0.96 (0.001)        & 1.83 (0.075)        & 0.02 (0.007)        & 0.30 (0.022)       \\
		Right lung              &&                    &                    &                    &                    && 0.94 (0.001)        & 1.60 (0.053)        & 0.00 (0.003)        & 0.23 (0.010)       \\
		Heart                   && 0.92 (0.003)       & 1.42 (0.046)       & -0.01 (0.006)      & 0.47 (0.015)       && 0.92 (0.002)        & 1.21 (0.039)        & -0.00 (0.009)       & 0.50 (0.016)       \\
		Spinal cord             && 0.75 (0.004)       & 1.06 (0.005)       & -0.18 (0.005)      & 0.38 (0.005)       && 0.72 (0.004)        & 0.98 (0.009)        & -0.14 (0.013)       & 0.39 (0.005)       \\
		\cmidrule{1-1} \cmidrule{3-6} \cmidrule{8-11}
		Left breast             && 0.89 (0.004)       & 3.52 (0.184)       & -0.02 (0.008)      & 1.71 (0.128)       && 0.86 (0.004)        & 3.93 (0.195)        & -0.01 (0.016)       & 1.67 (0.125)       \\
		Right breast            && 0.91 (0.003)       & 2.14 (0.046)       & 0.17 (0.156)       & 0.68 (0.016)       && 0.92 (0.003)        & 2.03 (0.048)        & 0.02 (0.010)        & 0.73 (0.020)       \\
		Left breast w.   ax     && 0.86 (0.005)       & 4.08 (0.115)       & 0.05 (0.010)       & 1.39 (0.042)       && 0.85 (0.005)        & 4.18 (0.113)        & 0.16 (0.093)        & 1.42 (0.046)       \\
		Right breast w.   ax    && 0.86 (0.004)       & 3.93 (0.086)       & 0.12 (0.076)       & 1.24 (0.028)       && 0.84 (0.004)        & 5.00 (0.124)        & 0.05 (0.009)        & 1.31 (0.031)       \\
		Left lung               && 0.97 (0.001)       & 1.53 (0.063)       & 0.00 (0.004)       & 0.29 (0.022)       && 0.97 (0.001)        & 1.64 (0.067)        & 0.02 (0.007)        & 0.30 (0.022)       \\
		Right lung              && 0.97 (0.001)       & 1.60 (0.055)       & -0.01 (0.002)      & 0.22 (0.007)       && 0.98 (0.001)        & 2.05 (0.069)        & 0.00 (0.004)        & 0.23 (0.009)       \\
		Heart                   &&                    &                    &                    &                    && 0.93 (0.002)        & 1.53 (0.048)        & -0.01 (0.008)       & 0.49 (0.015)       \\
		Spinal cord             && 0.75 (0.004)       & 1.09 (0.014)       & -0.16 (0.008)      & 0.39 (0.005)       && 0.75 (0.004)        & 1.06 (0.008)        & -0.16 (0.011)       & 0.40 (0.005)       \\
		\cmidrule{1-1} \cmidrule{3-6} \cmidrule{8-11}
		Left breast             && 0.89 (0.004)       & 3.82 (0.191)       & 0.01 (0.024)       & 1.74 (0.130)       && 0.88 (0.004)        & 3.75 (0.187)        & -0.03 (0.007)       & 1.70 (0.128)       \\
		Right breast            && 0.91 (0.003)       & 2.26 (0.054)       & 0.01 (0.005)       & 0.71 (0.018)       && 0.92 (0.003)        & 2.27 (0.062)        & 0.03 (0.012)        & 0.76 (0.023)       \\
		Left breast w.   ax     && 0.86 (0.005)       & 4.25 (0.115)       & 0.23 (0.157)       & 1.34 (0.041)       && 0.85 (0.005)        & 4.30 (0.118)        & 0.06 (0.021)        & 1.46 (0.051)       \\
		Right breast w.   ax    && 0.86 (0.004)       & 3.90 (0.087)       & 0.09 (0.054)       & 1.30 (0.031)       && 0.86 (0.004)        & 4.11 (0.100)        & 0.06 (0.011)        & 1.36 (0.036)       \\
		Left lung               && 0.97 (0.001)       & 1.50 (0.061)       & 0.01 (0.005)       & 0.29 (0.021)       && 0.97 (0.001)        & 1.51 (0.063)        & 0.02 (0.005)        & 0.30 (0.021)       \\
		Right lung              && 0.97 (0.001)       & 2.08 (0.088)       & -0.01 (0.004)      & 0.24 (0.010)       && 0.97 (0.001)        & 1.58 (0.053)        & 0.00 (0.004)        & 0.26 (0.015)       \\
		Heart                   && 0.93 (0.002)       & 1.43 (0.046)       & -0.01 (0.005)      & 0.51 (0.016)       && 0.94 (0.002)        & 1.42 (0.046)        & 0.02 (0.013)        & 0.51 (0.017)       \\
		Spinal cord             &&                    &                    &                    &                    && 0.75 (0.004)        & 1.11 (0.009)        & -0.16 (0.010)       & 0.42 (0.005)       \\		
        \bottomrule
    \end{tabular}
    \end{adjustbox}
    \caption{Mean \gls{dsc} (higher is better), \gls{hd95} (lower is better), \gls{ravd} (lower is better), and \gls{assd} (lower is better) and their \glspl{se} (in parenheses) computed on the test set before and after using incremental learning. The figures were reported on the eight evaluated structures: left breast, right breast, left breast with lymph nodes, right breast with lymph nodes, left lung, right lung, heart, and spinal cord.
    }
    \label{tab:incremental_0123}
\end{table*}

\section{Discussion} \label{sec:discussion}

In this section, we compare the proposed model to other recent approaches using all metrics introduced in \secref{subsec:eval}. We then discuss the performance on the task of incremental learning from $k-1$ to $k$ classes when employing the proposed \loss{}. Finally, we discuss the qualitative results of the baseline models, the \gls{ssl} model, and the proposed approach on the segmentation task.

\subsection{Quantitative Analysis}

The Nemenyi post-hoc test (\tableref{tab:result_nemenyi}) reveals that the proposed methods performed on par with the baseline models, and that the \gls{ssl} method performed significantly worse than the other methods.
In \tableref{tab:single} we see that: (i) the performance of the full model with the \loss{} is comparable to the eight single models in all evaluated metrics while reducing the training time by a factor of seven, (ii) the models with the two proposed voxel-based and slice-based \losspl{} did not outperform the baseline models, nor the model with the \loss{} without weights, and (iii) the proposed method outperformed the \gls{ssl} model by large margins in all the evaluated metrics, with a much shorter training time. A possible explanation as to why the \gls{ssl} model under-performs compared to the other methods is that we used the full set of labels containing both the available delineations and the predicted pseudo masks in the \gls{ssl}. It thus raised a problem that the pseudo masks, that were generated by the teacher models, might be wrong, making the student model solve the wrong problem. 

As can be seen in \tableref{tab:single}, the \gls{dsc} and \gls{hd95} scores of the spinal cord on all models tend to be the worst ($0.75$--$0.76$) and the best ($1.05$--$1.06$), respectively, compared to the other organs. These findings could be explained by the fact that compared to the other classes, the spinal cord occupies a much smaller area. This explanation is supported when comparing the last row of \figref{fig:qualitative} (small regions) to other rows (large regions). The proposed weighted models were intended to resolve this problem, but it turned out that the performance was unchanged whether or not we used either weighting scheme. This implies that the model with the proposed \loss{} is not sensitive to the amount of data available for each class, nor to the size of the structures in the classes.

From \tableref{tab:single}, we also see that when comparing the performance of the baseline models and the proposed model on the \gls{ravd} numbers, the baseline models seem to make over-segmentation on all classes (seven are positive and one is negative), while the proposed method appears to be more balanced (five are negative and three are positive). 

It can be seen from \tableref{tab:incremental_0123} that the evaluated metrics are similar for the $k-1$ initial classes before and after performing incremental learning using the proposed method with the \loss{}. This means that the knowledge on existing models was retained, and further transferred to the new class when new training data became available. In addition to that, comparing \tableref{tab:incremental_0123} and \tableref{tab:single}, we see that the new models trained on $k$ structures facilitating incremental learning perform on par with the baseline models for the added \glspl{oar}, implying that the models with the \losspl{} work well in the incremental learning setting.

By looking at the \figref{fig:learn_curve}, it is interesting to note that in all incremental learning experiments the \gls{dsc} of the additional \glspl{oar} converged very quickly (after being added in epoch 96) and the \gls{dsc} scores of the existing organs remained unchanged after the additional \glspl{oar} were added.

\subsection{Qualitative Analysis}

Predictions in \figref{fig:qualitative} might not be an accurate representation of the performance of the models since the slices are randomly selected. The baseline models and the proposed method are in many of the examples more similar to the delineations compared to the \gls{ssl} model. One example can be found in the first row where the predictions of the \gls{ssl} model are empty for both the left breast and the left breast with lymph nodes. Other examples are predictions of the right breast with lymph nodes in the second row as well as in the left breast with lymph nodes in the third row. In the randomly selected single slices in \figref{fig:qualitative}, the \gls{ssl} model under-predicted more often than the baseline models and the proposed method.

These observations align with the quantitative analysis and the results in \tableref{tab:incremental_0123}, where the \gls{ssl} model under-performs compared to the baseline models and the proposed method.

\section{Conclusion}

We have presented a novel \loss{} for semantic image segmentation. The proposed method has not only been shown to work well when training on incomplete data but also when compared to a state-of-the-art \gls{ssl} method. Furthermore, the proposed method works well in the incremental learning setting, where it is able to learn new structures without forgetting the ones that were already learned. Interesting venues for future work might include, for instance, how rapidly a new structure is learned, or how much data is required to learn a new structure.



\bibliographystyle{IEEEtran}
\bibliography{ref}

\end{document}

%% file: glossaries.tex
\makeglossaries

\newacronym[plural=CNNs,firstplural=Convolutional Neural Networks (CNNs)]{cnn}{CNN}{Convolutional Neural Network}

\newacronym[plural=FCNs,firstplural=Fully Convolutional Networks (FCNs)]{fcn}{FCN}{Fully Convolutional Network}

\newacronym{ml}{ML}{machine learning}
\newacronym{dl}{DL}{deep learning}
\newacronym{qcmlb}{QC-MLB}{Question-Centric Multimodal Low-rank Bilinear}
\newacronym{bert}{BERT}{Bidirectional Encoder Representations from Transformers}
\newacronym{bleu}{BLEU}{Bilingual Evaluation Understudy}
\newacronym{mlm}{MLM}{Masked Language Model}
\newacronym{nsp}{NSP}{Next Sentence Prediction}
\newacronym{relu}{ReLU}{rectified linear unit}
\newacronym{lrelu}{LeakyReLU}{leaky rectified linear unit}
\newacronym{nn}{NN}{neural network}
\newacronym{mri}{MRI}{magnetic resonance imaging}

\newacronym{ct}{CT}{computed tomography}
\newacronym{t1c}{T1c}{post-contrast T1-weighted}
\newacronym{t2}{T2w}{T2-weighted}
\newacronym{t1}{T1w}{T1-weighted}
\newacronym{flair}{FLAIR}{T2  Fluid  Attenuated  Inversion  Recovery}
\newacronym{lgg}{LGG}{low-grade glioma}
\newacronym{hgg}{HGG}{high-grade glioma}

\newacronym{svm}{SVMs}{Support-vector Machines}
\newacronym{crf}{CRF}{Conditional Random Field}
\newacronym{vae}{VAE}{Variational Auto-Encoder}
\newacronym{tunet}{TuNet}{End-to-end Hierarchical Tumor Segmentation using Cascaded Networks}
\newacronym{dram}{DRAM}{dynamic random access memory}
\newacronym[plural=GPUs,firstplural=graphics processing units (GPUs)]{gpu}{GPU}{graphics processing unit}

\newacronym{hpc2n}{HPC2N}{High Performance Computer Center North}
\newacronym{snic}{SNIC}{Swedish National Infrastructure for Computing}

\newacronym{dsc}{DSC}{S{\o}rensen-Dice coefficient}
\newacronym{se}{SE}{standard error}
\newacronym{seb}{SEB}{Squeeze-and-Excitation block}
\newacronym[plural=REBs,firstplural=ResNet blocks]{res}{REB}{ResNet block}
\newacronym{hd}{HD}{Hausdorff distance}
\newacronym{hd95}{HD95}{$95^{th}$ percentile of the Hausdorff distance}
\newacronym{dauc}{DAUC}{Dice Area Under Curve}
\newacronym{rftp}{RFTPs}{Ratio of Filtered True Positives}
\newacronym{rftn}{RFTNs}{Ratio of Filtered True Negatives}

\newacronym[plural=SDs,firstplural=standard deviations (SDs)]{sd}{SD}{standard deviation}
\newacronym{brats}{BraTS 2020}{Brain Tumors in Multimodal Magnetic Resonance Imaging Challenge 2020}

\newacronym{mdnet}{MDNet}{End-to-end Multi-Decoder Cascaded Network for Tumor Segmentation}
\newacronym{ce}{CE}{categorical cross-entropy}

\newacronym{miccai}{MICCAI}{International Conference on Medical Image Computing and Computer Assisted Intervention}

\newacronym[plural=OARs,firstplural=organs-at-risks (OARs)]{oar}{OAR}{organs-at-risk}

\newacronym{loss}{loss}{Data-Adaptive Loss Function}

\newacronym{ravd}{RAVD}{relative absolute volume difference}
\newacronym{assd}{ASSD}{average symmetric surface distance}

\newacronym{dna}{DNA}{deoxyribonucleic acid}

\newacronym{asd}{ASD}{symmetric surface distance}
\newacronym{tpe}{TPE}{Tree-structured Parzen Estimator Approach}
\newacronym{rmse}{RMSE}{root mean square error}
\newacronym{ssl}{SSL}{semi-supervised learning}